\documentclass[aps,prb,twocolumn,groupedaddress]{revtex4-1}
\usepackage{graphicx}
\usepackage[caption=false]{subfig}
\usepackage{float}
\usepackage{amsmath}
\usepackage{color}
\usepackage[dvipsnames]{xcolor}

\usepackage[english]{babel}
\usepackage{url}
\usepackage{hyperref}
\usepackage{ae,aecompl}
\begin{document}

\raggedbottom   %to avoid unwanted extra vertical space

\newcommand{\ox}{Al$_2$O$_3$}
\newcommand{\wad}{$W_{\text{Adh}}(d)$}
\newcommand{\duber}{$d_{0_{\text{UBER}}}$}
\newcommand{\dspline}{$d_{0_{\text{spline}}}$}
\newcommand{\wuber}{$W_{0_{\text{UBER}}}$}
\newcommand{\wspline}{$W_{0_{\text{spline}}}$}

\title{Structural details of Al/\ox~junctions and their role \\ in formation of electron tunnel barriers}

\author{M. Koberidze} %\thanks{A.A@university.edu}}
	\affiliation{COMP Center of Excellence, Department of Applied Physics,  Aalto University School of Science, P.O. Box 11100, FI-00076 Aalto, Finland}
\author{M. J. Puska}
	\affiliation{COMP Center of Excellence, Department of Applied Physics,  Aalto University School of Science, P.O. Box 11100, FI-00076 Aalto, Finland}
\author{R. M. Nieminen}
	\affiliation{COMP Center of Excellence, Department of Applied Physics,  Aalto University School of Science, P.O. Box 11100, FI-00076 Aalto, Finland}

\begin{abstract}

We present a computational study of the adhesive and structural properties of the Al/\ox~interfaces as building blocks of the Metal-Insulator-Metal (MIM) tunnel devices, where electron transport is accomplished via tunnelling mechanism through the sandwiched insulating barrier. The main goal of this paper is to understand, on the atomic scale, the role of the geometrical details in the formation of the tunnel barrier profiles. To provide reliable results, we carefully assess the accuracy of the traditional methods used to examine Al/\ox~interfaces. These are the most widely employed exchange-correlation functionals, LDA, PBE and PW91, the Universal Binding Energy Relation (UBER) for predicting equilibrium interfacial distances and adhesion energies, and the ideal work of separation as a measure of junction stability. Finally, we perform a detailed analysis of the atomic and interplanar relaxations in each junction. Our results imply that the structural irregularities on the surface of the Al film have a significant contribution to lowering the tunnel barrier height, while interplanar relaxations in the Al film, away from the immediate interface do not have a notable impact on the tunnelling properties. On the other hand, up to 5-7 layers of \ox~may be involved in shaping the tunnel barriers. Interplanar relaxations of these layers depend on the geometry of the interface and may result in the net contraction by 13\% relative to the corresponding thickness in the bulk oxide. This is a significant amount as the tunnelling probability depends exponentially on the barrier width.

\end{abstract}

\maketitle
\section{Introduction}

Metal-Insulator-Metal (MIM) tunnel junctions are principal components of numerous modern nanoelectronic devices. The operation of such devices is based on electron tunnelling across MIM systems from one metal to the other. The most common choice for the metal is Al, and for the insulator, its native oxide, \ox. Thus, Al/\ox~junction is the most widely used base structure in MIM tunnelling devices, particularly, for the applications such as quantum computing, ultra-sensitive magnetometry, radiation detection, quantum metrology\cite{Seidel2015, Narlikar2017}.

It has been shown that the electron transport properties of Al/\ox-based tunnel devices are strongly affected by the atomic structure near the interface between Al and \ox \cite{JungDFT, Alimardani2012, Zeng2016, Koberidze2016}, as well as by the variation in the thickness of the oxide layer throughout the junction \cite{Ville2011, Greibe2011, Zeng2015}. Structural defects at the interface \cite{JungDFT,  Zeng2016}, roughness of the metal substrate \cite{Kohlstedt1996, Alimardani2012, Conley2013} and non-uniformity of the oxide thickness \cite{JungDFT, Ville2011, Greibe2011} are anticipated to be the critical factors, which may deteriorate the functionality of the oxide as a barrier. However, characterization of this impact is in its early stage and hence uncertain yet. Understanding the subtle relationship between the geometry of the Al/\ox~junctions and the tunnelling properties on the atomistic level would help to achieve a better control and an improved performance of MIM devices. Therefore, careful atomistic characterization of the interface is essential.

Although the structure of Al/\ox~systems has actively been explored over the decades both theoretically \cite{Zhang2000, Batyrev2001, Siegel2002, Dieskova2007, Kim2013, Liu2014, Pilania2014, Dubois2016} and experimentally \cite{ Timsit1985, Medlin1997, Toofan1998, Dehm2002,  Alimardani2012, Flototto2015, Zeng2015, Zeng2016}, an in-depth understanding of the interface properties is still an ongoing issue. Experimental observations have provided valuable information, for example, about the most probable crystallographic orientation relationship between Al and \ox~at the interface \cite{Medlin1997, Dehm2002}, about Al substrate roughness  \cite{Timsit1985, Alimardani2012}, chemical state of the ions \cite{Jeurgens2002}, oxide thickness distribution \cite{Zeng2015}, bond lengths and coordinations of the atoms \cite{Zeng2016} at the interface, Al/O ratios for different oxide thicknesses \cite{Flototto2015} and for different oxidation times and temperatures \cite{Jeurgens2002}. However, experimental determination of the detailed atomic structures of the buried, ultrathin interfaces is challenging. Besides, the structure of the formed interface depends on the oxidation method and on the parameters used during the oxidation \cite{Bianconi1979, Jeurgens2002, Zeng2015}. This further complicates obtaining consistent experimental data about the atomic details of the interface geometry and their effects on the tunnelling properties. For these reasons, availability of accurate and reliable theoretical predictions is particularly important. 

Despite the significance of the Al/\ox~junctions, and popularity of the Density Functional Theory (DFT) method for quantum mechanical description of the many-body systems, the first DFT studies on Al/\ox~interfaces appeared relatively late. Pioneering DFT works addressed the atomic structure of differently constructed Al/\ox~interfaces, bonding at the interface, adhesion energies and the most stable terminations of the oxide at the interface \cite{Zhang2000, Watari2000, Batyrev2001, Siegel2002, Kiejna2002}. More recent DFT studies are focused on improving junction models and on screening possibilities for improved adhesion, which is the measure of the structural stability of the system \cite{Kim2013, Liu2014, Dubois2016}.

Nevertheless, how the structure and geometry of the Al/\ox~interfaces relate to the tunnelling properties of Al/\ox-based devices, is barely known. In our previous study \cite{Koberidze2016}, we demonstrated the effect of the different interface structures on the tunnel barrier heights and widths for the Al/{\ox}/Al tunnel junctions. The work was based on the analysis of experimental current-voltage (IV) data using classical and semi-classical models for tunnelling current, where the employed barrier parameters were those predicted by DFT. In the current paper, first, we explain the procedure for obtaining the model Al/\ox~junctions employed in our preceding work. Next, we present the detailed structural analyses of the model systems to understand the role of atomic-scale geometrical variations in the formation of tunnel barriers. Particularly, we identify the factors affecting the height of the tunnel barrier, and those contributing to the thickness variation of the oxide. To our knowledge, none of the previous theoretical studies have addressed the structural details of the Al/\ox~junctions in connection with the tunnel barrier profiles, neither factors contributing to the non-uniformity of the oxide barrier have been investigated theoretically.

Since tunnelling properties of the MIM systems are sensitive to the interface geometry, when modelling tunnel junctions, obtaining reliable optimal structures and performing careful analysis are critical. Therefore, we pay particular attention to validating the applied methods. One of the preconditions for accurate DFT modelling is choosing a suitable exchange-correlation functional. As we are going to simulate interfaces, the employed functional must be able to properly describe both Al and \ox. We  compare the accuracies of the most widely used standard functionals in predicting the bulk and surface properties of individual materials.

Another important issue is an accurate determination of the adhesion energies as they are used in evaluating the stabilities of the junctions, and are linked to the bonding at the interface. Therefore, we assess the accuracy of the Universal Binding Energy Relation (UBER) which is commonly adopted for determining equilibrium interfacial distances and equilibrium adhesion energies. In addition, we challenge the reliability of conventional interpretation of the computed ideal work of adhesion as a measure of the interface stability. 

Having examined the performance of the different methods and identified the most accurate ones for our systems, we construct the optimal model interfaces. In the end, we look into the correlation between the geometrical structure and the tunnel barrier profile. For this purpose, we analyse the atomic, as well as interplanar relaxations. Atomic positions at the interface play an important role in defining the magnitude of tunnel barrier heights. This is because together with the charge transfer, they contribute to shaping the charge distribution at the interface. The charge distribution, in turn, affects electron energy level alignment in the junction which determines the height of the tunnel barrier. In addition to atomic relaxations at the interface, interplanar relaxations beyond the immediate interface contribute to the thickness variation of the oxide, thus, affecting the width of the tunnel barrier. Even a tiny change in the width of the barrier significantly impacts on the functionality of the MIM tunnel devices, since the tunnelling probability depends exponentially on the barrier width.  

The present article is organized as follows: In section \ref{Calcmethods}, we describe our calculation set-up. In section \ref{bulks}, we  compare the different exchange-correlation functionals for describing the properties of bulk Al and \ox. We examine the accuracy of the same functionals for predicting the surface properties of Al and \ox~slabs in section \ref{surfaces}.
In the first part of section \ref{sinterfaces}, we describe the procedure for setting up the model interfaces and a step-by-step validation of the applied methods. The second part of section \ref{sinterfaces} includes analysis of the atomic and interplanar relaxations and their connections to the tunnel barrier profiles.

\section{Calculation method}\label{Calcmethods}

In all the calculations presented in this paper, we use the density-functional theory (DFT) within the projector-augmented wave code GPAW \cite{GPAW, GPAWrev}. In our set-up, electron occupation function is represented with the Fermi-Dirac distribution at the 0.1 eV electronic temperature.

Until recently, the most frequently used exchange-correlation functionals for Al/Al$_2$O$_3$ interfaces have been the local density approximation (LDA \cite{Siegel2002, Watari2000}) and the generalized gradient approximations (GGAs) PW91 \cite{Siegel2002, Kim2013} and PBE \cite{Liu2014, JungDFT, Dieskova2007}. To find the most suitable one for our calculations, we compare the accuracy of the three density functionals in predicting equilibrium lattice constants ($a_0$), cohesive energies ($E_c$) and bulk moduli ($B$) of the bulk materials, as well as interplanar space  relaxations ($\Delta d$) and surface energies ($\sigma$) of the slabs. Calculation details specific to the individual studied systems are mentioned in the corresponding sections.

\section{Bulk A\lowercase{l} and $\alpha$-A\lowercase{l}$_2$O$_3$} \label{bulks}

In our set-up, as a starting point, both the Al and \ox ~unit cell dimensions are set corresponding to the experimental lattice constants, for the cubic Al, $a_0$=4.05 \AA \cite{Wyckoff1963} and for the hexagonal \ox, $a_0$=4.759 \AA, $c_0$=12.991 \AA \cite{Newnham1962}. In our calculations,  bulk Al is represented with a hexagonal unit cell containing 12 atoms. To sample the Brillouin zone, we use the (14x14x14) Monkhorst-Pack grid which corresponds to 1372 k-points in the irreducible part of the Brillouin zone. The real-space grid spacing is set to 0.18 \AA. For the bulk $\alpha$-\ox, we use a hexagonal unit cell as well, containing 30 atoms, among them 12 Al and 18 O. We employ the (4x4x4) Monkhorst-Pack grid, i.e., 32 k-points in the irreducible part of the Brillouin zone and the 0.13 \AA~ real-space grid spacing. With these parameters, the total energy per atom for both the materials is converged to within 1 meV. We find the equilibrium lattice constants, bulk moduli and cohesive energies from the equation of state (EOS). For this purpose, we strain the initial volumes isotropically by factors between 0.95 and 1.05 and calculate energy-versus-volume curves using 10 equally-spaced points. We use the Murnaghan function \cite{Murnaghan1944} to fit to the calculated points. The resulted bulk and surface parameters are listed in table \ref{XC_test}. Errors with respect to the known experimental data are also provided. Since experimental measurements are performed at finite temperatures, values extrapolated to 0K are used where available. 

\begin{table*}
\caption{Equilibrium lattice constants ($a_0$), bulk moduli ($B$) and cohesive energies ($E_c$) of bulk Al and \ox. The error percentages are given with respect to the experimental values listed on the last row.} 
\begin{center}
\begin{tabular}{@{}*{16}{c}}
\hline\hline
\multicolumn{1}{c}{} & \multicolumn{6}{c}{Al} & \multicolumn{8}{c}{Al$_2$O$_3$} \\
\cline{2-7} 
\cline{9-16}
~ & $a_0$ ~ & Error ~  & $B$~  & Error ~ & $E_c$~ & Error & ~ & ~ $a_0$ ~ & Error ~ & ~ $c_0$ ~ & Error  & $B$~  & Error ~ & $E_c$~ & Error \\
~ & (\AA) ~ & (\%) ~  & (GPa)~  & (\%) ~ & (eV/atom)~ & (\%) & ~ & ~ (\AA) ~ & (\%) ~ & ~ (\AA) ~ & (\%)  & (GPa)~  & (\%) ~ & (eV/Al$_2$O$_3$)~ & (\%) \\
\hline
LDA & 3.99 & -1.48 & 84.1 & 6.01 & 4.00 & 17.99 & ~ & ~ 4.749 & -0.21 & 12.967 & -0.18 & 258.2 & 1.57 & 36.35 & 14.31 \\
PBE & 4.04 & -0.25 & 77.9 & -1.86 & 3.43 & 1.18 & ~ & ~ 4.829 & 1.47 & 13.184 & 1.49 & 230.2 & -9.44 & 30.59 & -3.81 \\
PW91 & 4.05 & 0.0 & 74.4 & -6.27 & 3.38 & -0.29 & ~ & ~ 4.821 & 1.30 & 13.164 & 1.33 & 234.0 & -7.95 & 31.27 & -1.67 
\\
\\
experiment &~~ 4.05\cite{Wyckoff1963} & & 79.38 \cite{Kamm1964}  & & 3.39 \cite{Kittel7th} & ~ & ~ & ~ 4.759 \cite{	Newnham1962} & & 12.991 \cite{Newnham1962} & & 254.2 \cite{Gieske1968} & & 31.8 \cite{CRC} &  \\
\hline\hline
\end{tabular}
\end{center}
\label{XC_test}
\end{table*}
Among the three functionals, PW91 gives the most accurate results for the Al lattice constant and the cohesive energy  with respect to the experimental values. LDA gives the most precise lattice constants for \ox~and predicts its bulk modulus significantly better than the two gradient-corrected functionals. On the other hand, LDA results in an order of magnitude larger errors for the cohesive energies of both the materials compared to those predicted by PBE and PW91. Since we are going to estimate adhesion energies, the cohesive energy $E_c$ is an important parameter in our study. Taking the sum of the absolute values of all the errors, PW91 slightly outperforms PBE. The sums amount to 19.5 \%, 18.81 \%, and 41.75 \% for PBE, PW91, and LDA, respectively. However, taking into account the scatter in experimental values reported in literature, we should acknowledge that the less than 1 \% difference between the PW91 and PBE results does not necessarily mean the PW91 superiority over PBE.

A satisfactory description of bulk properties does not guarantee an adequate description of surfaces. Correct estimates for the surface energy and atomic relaxation are crucial to our interface calculations. We present surface analyses in the following section.

\section{A\lowercase{l}(111) and A\lowercase{l}$_2$O$_3$(0001) surfaces}\label{surfaces}

When starting the structure optimization of the Al and \ox~surfaces, we use the same experimental lattice constants as for setting up the bulk materials described in the previous section. We use the 2D-periodic slab model, which is non-periodic in the direction perpendicular to the surface (along the $z$ axis), and is periodic in the $xy$ plane. We add  5 \AA~vacuum on both surfaces of the slabs. 

We carry out calculations with the (4x4x1) Monkhorst-Pack grid for the Al metal and the (14x14x1) grid for the \ox~oxide. The real space grid spacings are the same as in the bulk calculations. We study the Al(111) and \ox(0001) surface orientations since in this way, the crystallographic structures of Al and \ox~are best matched, and they will also represent the constituents of the junctions in our forthcoming calculations. The same preferred orientations have also been verified in experiments \cite{Medlin1997}. In our simulations, the Al(111) and \ox(0001) slabs are composed of 5 and 18 layers,  respectively. It has been shown previously that 5 layers of Al(111) and 15 layers of \ox(0001) are enough to converge the surface energies and the surface relaxations \cite{Siegel2002}. However, we have represented the oxide with its complete hexagonal unit cell with 18 layers, which would be sufficient for recovering the ordered structure. This is relevant to our calculations, since we are interested in the geometry of the whole oxide film, at the interface and beyond it. Each layer of the Al slab contains 3 atoms, resulting in 15 atoms in total. The oxide slab is composed of 6 units of Al$_2$O$_3$, i.e., 30 atoms in total. 

\begin{figure}[h!]
	\subfloat[]{	
		\includegraphics[width=0.4\textwidth]{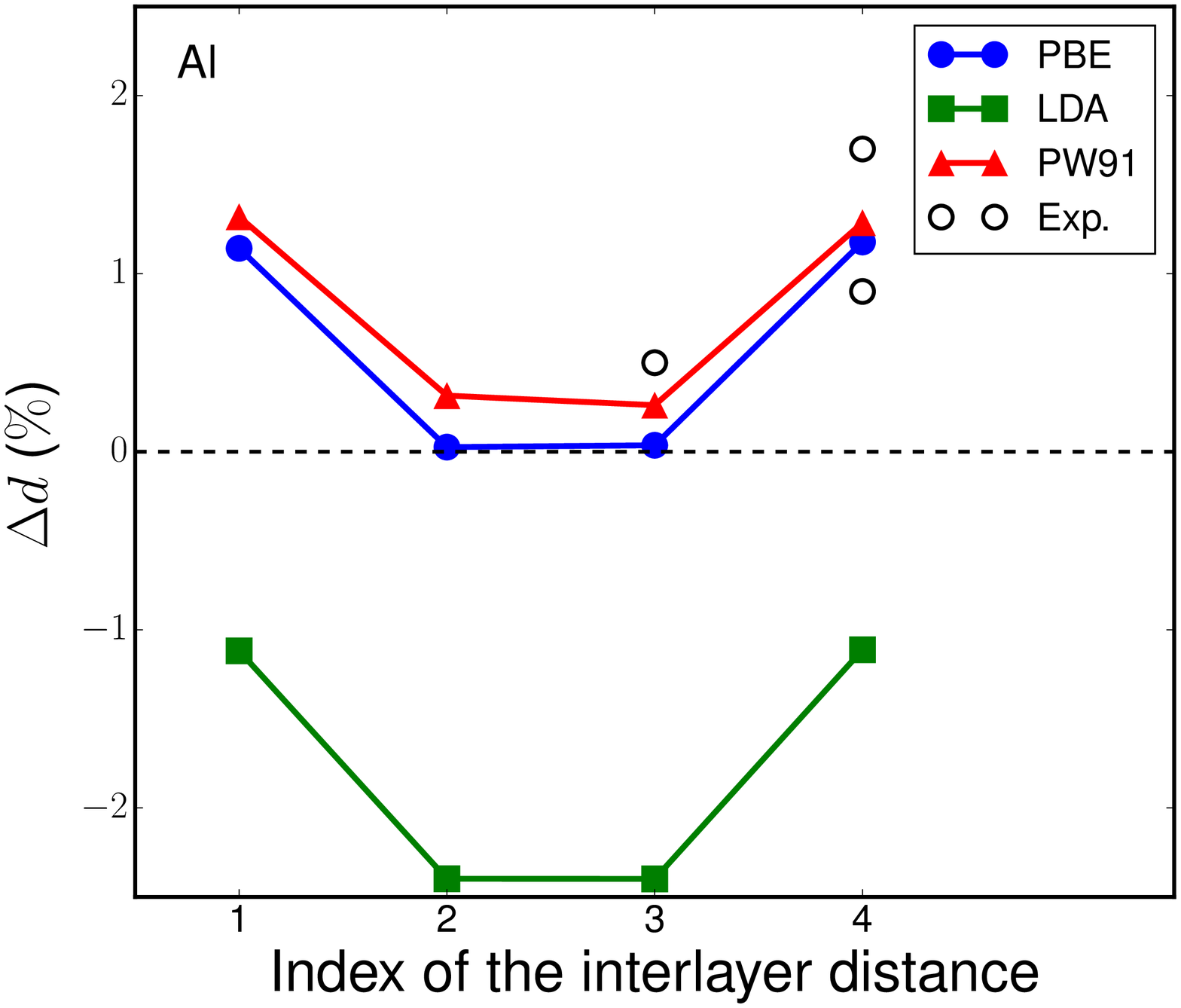} 
	}	\\
	\subfloat[]{	
		\includegraphics[width=0.4\textwidth]{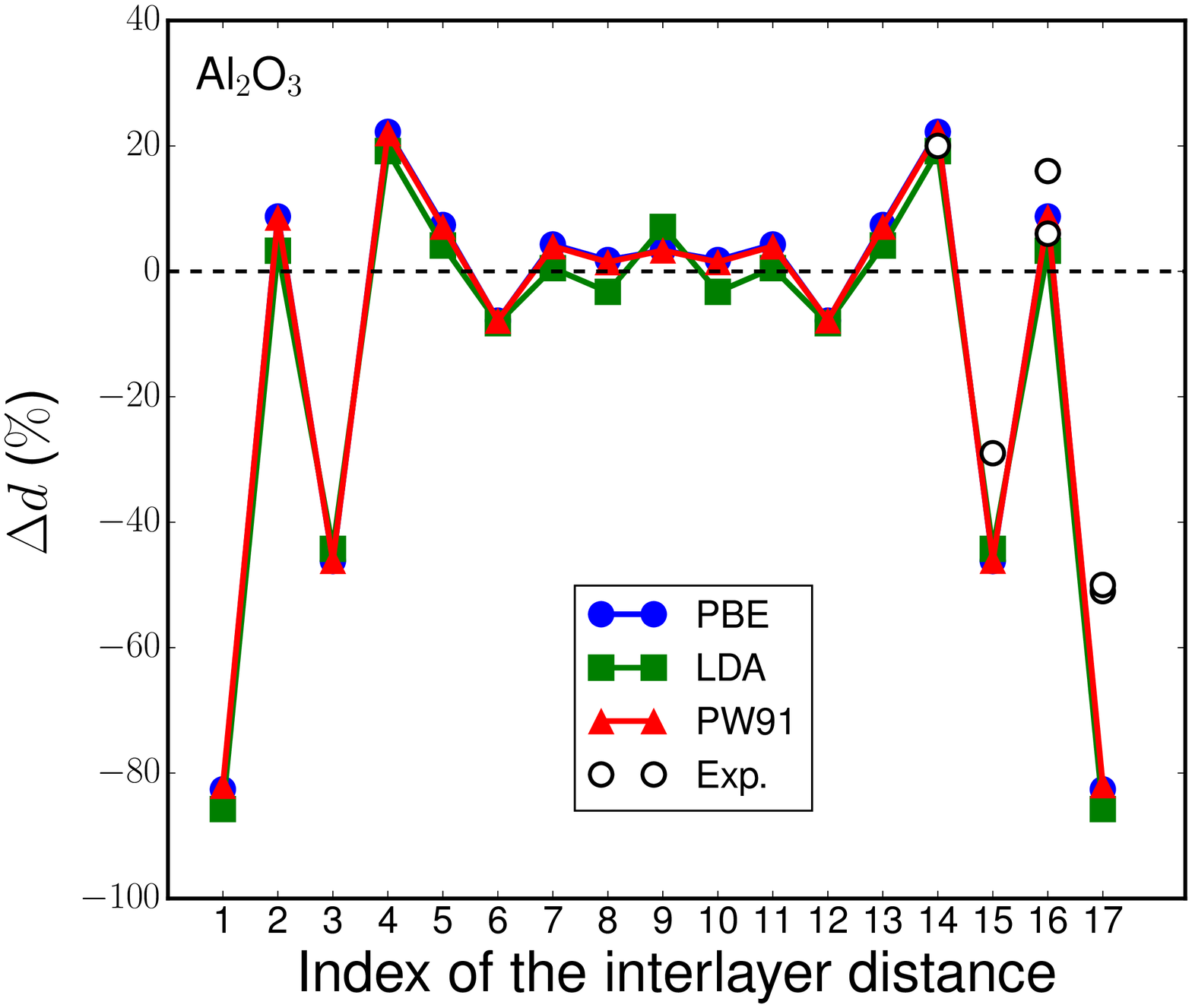}
	} 	
\caption{Surface relaxation of the (a) Al(111) and (b) Al-terminated \ox(0001) slabs calculated using the LDA, PBE and PW91 exchange-correlation functionals. The vertical axis shows the percentage change in the interplanar spacing after relaxation  $\Delta d=(d_{ij}-d_{0_{ij}})/d_{0_{ij}}\times 100$, where $d_{ij}$ is the distance between the successive layers $i$ and $j$ of the relaxed slab. $d_{0_{ij}}$ is the same quantity for the unrelaxed slab. The horizontal axis labels the index of $\Delta d$. For example, the index 1 means the first interlayer distance, i.e., the distance between the first and the second bottom layers. The white circles represent experimental data: for Al, 1.7$\pm$ 0.3 \% \cite{Noonan1990} and 0.9 $\pm$ 0.5 \% \cite{Nielsen1982} for the topmost interlayer distance (index 4), and 0.5 $\pm$ 0.7\%  \cite{Noonan1990} for the preceding interlayer distance (index 3). For \ox,~ the corresponding values are -51 \% \cite{Guenard1998} and -50 \% \cite{Soares2002} for the topmost interlayer distance (index 17), 16 \% \cite{Guenard1998} and 6 \% \cite{Soares2002} for the second topmost interlayer distance (index 16), and -29 \% \cite{Guenard1998} and 20 \% \cite{Guenard1998} for the two preceding interlayer distances (indexes 15 and 14, respectively).
}
\label{surfrelax}
\end{figure}

Keeping in mind that our slab structures will compose the interface later, when setting the lateral lattice constants, care must be taken that the lattice mismatch between Al and \ox~is consistent with the experimental findings. One option is to use the equilibrium lateral lattice constants predicted by the chosen exchange-correlation functional for the bulk systems. However, based on our calculations, the lattice mismatch $(a_0$(Al)-$a_0$(Al$_2$O$_3$))/$a_0$(Al) suggested by the PBE functional is 2.4 \%, while the experimentally evaluated mismatch amounts to 4.3 \% (\cite{Medlin1997}). We assume that maintaining the lattice mismatch close to the experimental value, will yield more realistic description of the junction. Therefore, during the relaxation of the surfaces, we fix the lateral lattice constants at the experimental values. Vertical relaxation is unconstrained and facilitated by the presence of the vacuum layer. The structures are relaxed until forces on each atom are less than 0.01 eV/\AA. Percentage changes in interplanar spacings $\Delta d$ resulting from the surface relaxation are illustrated in figure \ref{surfrelax}. 

For Al, both PBE and PW91 predict a slight expansion of interlayer spacings and produce results within the range of experimental errors, while LDA suggests relaxation with the negative sign contrary to the experimental observations (Figure \ref{surfrelax}(a)). This is also the case for other metals when experimental lattice constants are used \cite{DaSilva2006}. The correct atomic relaxation of the surface is the key ingredient for our junction calculations, to ensure the proper structure formation of different interfaces. Relaxation of the oxide is more complex and involves not only vertical displacements of the atomic planes but also expansion and rotation of the subsurface oxygen triangle. The rotation amounts to 2.9$^0$ with respect to the Al atom, around which the O$_3$ triangle is centred. The O-O bonds in the triangle are expanded by 2.6 \% compared to those in the bulk. The same effect has been observed in earlier theoretical and experimental works \cite{Batyrev2001, Soares2002, Toofan1998}. 

Experimental data for the changes in interlayer separation in \ox~exhibit a large scatter, not only with respect to the magnitudes but also to the sign \cite{Guenard1998, Soares2002, Toofan1998}. In figure \ref{surfrelax} (b), the experimental values are taken from references \cite{Guenard1998, Soares2002}. All the three functionals describe the oxide surface relaxation equally reasonably. 

Surface energies of Al and \ox~slabs are given in table \ref{surfE}. Our calculated surface energies are in good agreement with other theoretical works. However, to our  knowledge, there is a lack of experimental data for the Al(111) and \ox(0001)~ surface energies. In addition, they differ notably for \ox. This makes it difficult to evaluate the accuracy of the exchange-correlation functionals in predicting surface energies. Taking an average experimental value for \ox~as a reference, LDA tends to be in a better agreement with the experiments. PBE and PW91 predict lower surface energies than LDA by 12-17\%. Out of the two GGAs, PBE results in a smaller total error.

In summary, among the three functionals, the biggest advantage of LDA is to best predict the bulk properties of the oxide with the exception of the cohesive energy. On the other hand, PBE and PW91 provide a significantly better description of the bulk properties and surface relaxations of Al as well as the cohesive energies of both the materials compared to LDA. At the same time, they produce acceptable errors for the oxide properties best described with LDA. 
Therefore, we exclude LDA out of the three functionals. The two GGAs do not exhibit major differences in their performances, however, on average, PBE predicts the properties of Al slightly more accurately than PW91. Therefore,  we choose to continue calculations with the PBE functional.

\begin{table}[h]
\caption{Surface energies in J/m$^2$ for Al(111) and Al-terminated \ox(0001)~ slabs calculated with the LDA, PBE and PW91 exchange-correlation functionals. The values without references represent our work.}
\begin{center}
\begin{tabular}{@{}*{8}{llllllll}}
\hline\hline
\multicolumn{1}{c}{} & \multicolumn{3}{c}{Al} & ~ & \multicolumn{3}{c}{Al$_2$O$_3$} \\
\cline{2-4} 
\cline{6-8}
LDA & 0.98 & 1.02$^a$ & 0.88$^b$ & & 2.01 & 2.12$^a$ & 1.98$^c$ \\
PBE & 0.84 & 0.77$^b$ & 0.81$^d$ & & 1.64 & 1.54$^d$ &  \\
PW91 & 0.79 & 0.81$^a$ &  & & 1.66 & 1.54$^e$ & 1.59$^a$ \\
\\
experiment & 1.14$^f$ &  & & & 1.69$^g$ & 2.6$^h$\\
\hline\hline
$a$-Reference \cite{Siegel2002} \\
$b$-Reference \cite{DaSilva2006} \\
$c$-Reference \cite{Felice1999} \\ 
$d$-Reference \cite{Liu2014} \\
$e$-Reference \cite{Pinto2004} \\
$f$-Reference \cite{Tyson1977} \\
$g$-Reference \cite{McHale1997} \\
$h$-Reference \cite{Navrotsky2003}\\

\end{tabular}
\end{center}
\label{surfE}
\end{table}

\section{A\lowercase{l}/A\lowercase{l}$_2$O$_3$ interfaces}\label{sinterfaces}
When setting up different interfaces, we follow the procedure presented in reference \cite{Siegel2002}. We start constructing the interfaces with the same 5-layer Al(111) and the 18-layer \ox(0001) slabs, which we used as starting structures for the surface relaxation tests in the previous section. Thus, neither of the slabs is relaxed beforehand and the initial lattice constants correspond to the experimental values. However, due to the difference in the lattice constants of the two materials, one of them should be strained. Since Al has a smaller bulk modulus and is softer than Al$_2$O$_3$, we compress Al in the $xy$ plane to match the lattice of the oxide, as often done in other works as well \cite{Siegel2002, Batyrev2001, Liu2014}. This means that the lateral lattice constant $a$ of our model junction corresponds to the experimental lattice constant of Al$_2$O$_3$, 4.759 \AA.

\begin{figure}{}
\includegraphics[width=0.50\textwidth]{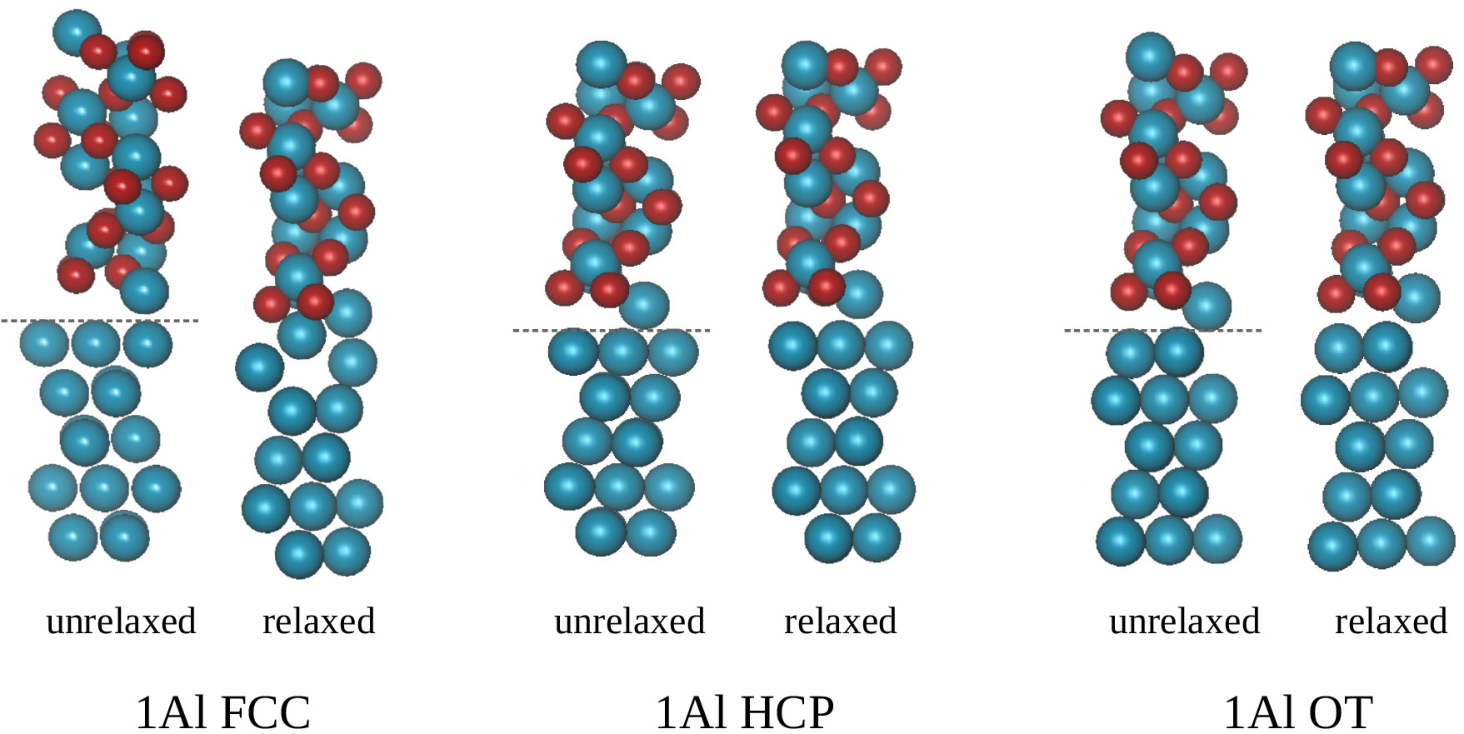} \\
\includegraphics[trim={0 0.0cm 0 0}, clip, width=0.50\textwidth]{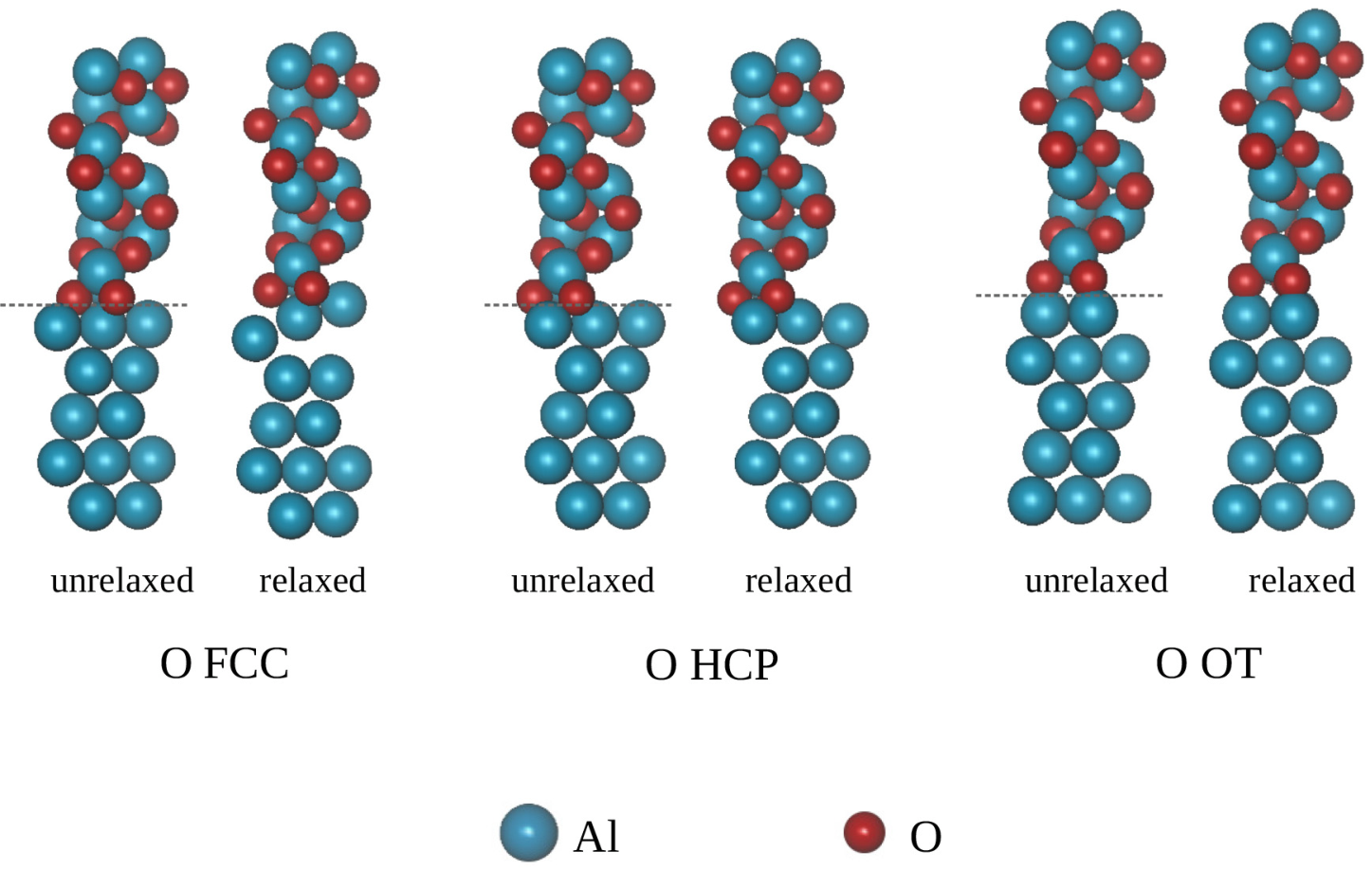}
\caption{The six modelled interfaces. Upper panel - the junctions with the initially Al-terminated oxides and three different stacking sequences at the interfaces. Lower panel - the junctions with the initially O-terminated oxides and with the same stacking sequences as in the Al-terminated structures. The dashed lines on the unrelaxed systems show the boundary between the metal and the oxide slabs separated by the distance $d_0$. The procedure for finding $d_0$ is explained in the text.}
\label{interfaces}
\end{figure}

Since it is impossible to study the whole configurational space for adhesion between two surfaces, we consider three different stacking sequences between the facing Al and \ox~surface layers, and two possible terminations of the oxide (a single Al or O layer), i.e., six configurations in total. For the stacking sequence we examine the face-centred (FCC), hexagonal-close-packed (HCP) and the octahedral (OT) sites. The FCC stacking labels the interface, where the Al surface atoms of the metal and those of the oxide sit on top of each other (see unrelaxed structures in figure \ref{interfaces}), the HCP stacking means that the surface Al atoms are placed along the second O layer of the oxide, and in the OT stacking the Al atoms of the metal sit on top of the first O layer of the oxide. In figure \ref{interfaces}, the $z$ axis is chosen perpendicularly to the interface. The whole system consists of 45 atoms for all geometries. 
 
All the calculations on interfaces are performed with the (4x4x1) k-point grid and a real space grid spacing of 0.13 \AA. A 5 {\AA}-vacuum layer is added on each side of the slab. We employ an asymmetric model, where the system is non-periodic in the $z$ direction. Thus, periodic boundary conditions are applied only in the $xy$ plane. To eliminate an artificial electric field due to the asymmetry of the slab, we apply the dipole correction to the electrostatic potential along $z$ axis as implemented in GPAW. We have chosen such construction of the interface over the periodic supercell model because presence of the vacuum layer in asymmetric model, allows full vertical relaxation of the structure, which is expected due to the lateral strain on Al and possible interfacial stresses. Conversely, freedom of translation in the $z$ direction is restricted in the supercell approach by the dimensions of the simulation box, where the vacuum layer is absent. Besides, asymmetric structure more closely resembles the Al/\ox~junction in experiments where the interface is indeed prone to relaxation before depositing the second electrode, while in the supercell model, Al/\ox~stacks are assumed to form an infinite array. Thus, our choice of the set-up enables us to  characterise the interface without imposing restrictions to the interplanar relaxations and without the influence of the second electrode. Advantages and disadvantages of the two methods are discussed, for example, in reference \cite[Chapter 5.2.2]{Oktyabrsky2010}.

\subsection{Adhesion and separation at the interface} 

\begin{table*}[]
\caption{Comparison of UBER and 5th order spline fits for finding the optimal distance $d_0$ between the unrelaxed slabs. Errors shown in bold are evaluated with respect to the DFT values. $W_\text{DFT}$(\duber) and $W_\text{DFT}$(\dspline) are adhesion energies calculated with DFT at $d_0$ predicted by UBER and the spline fits, respectively. Distances are given in \AA, adhesion energies in J/m$^2$ and errors in \%. }
\begin{tabular}{@{}*{12}{lccccccccccc}}
\hline
\hline
~ & \duber & \wuber & ~ & $W_\text{DFT}$(\duber) & \bf{Error}$_\text{UBER}$ & ~ & \dspline & \wspline & ~ & $W_\text{DFT}$(\dspline) & \bf{Error}$_\text{spline}$\\
%~ & (\AA) & (J/m$^2$) & ~ & (J/m$^2$) & ~ & (\AA) & (J/m$^2$) & ~ & (J/m$^2$) \\
\hline
1Al FCC & 2.66 & 1.13 & ~ & 1.04 & \bf{8.65} & ~ & 2.59 & 1.05 & ~ & 1.05 & \bf{0.00}\\
1Al HCP & 2.47 & 1.13 & ~ & 1.12 & \bf{0.89} & ~ & 2.40 & 1.14 & ~ & 1.13 & \bf{0.88}\\
1Al OT & 2.30 & 1.26 & ~ & 1.29 & \bf{-2.32} & ~ & 2.24 & 1.30 & ~ & 1.30 & \bf{0.00}\\
\\
O FCC & 1.41 & 6.95 & ~ & 6.61 & \bf{5.14} & ~ & 1.47 & 6.84 & ~ & 6.78 & \bf{0.88}\\
O HCP & 1.34 & 7.52 & ~ & 6.91 & \bf{8.83} & ~ & 1.40 & 7.31 & ~ & 7.09 & \bf{3.10}\\
O OT & 1.74 & 7.46 & ~ & 7.41 & \bf{0.67} & ~ & 1.73 & 7.41 & ~ & 7.41 & \bf{0.00}\\

\hline\hline

\end{tabular}

\label{fits}
\end{table*}
\begin{table*}
\caption{Effect of relaxation on $W_0$ and $d_0$. Arrows up indicate increase after relaxation, arrows down indicate decrease, $\sim$ describe almost unchanged value. Distances are given in \AA, adhesion energies in J/m$^2$. $d_{0 \text{(LDA)}}$ and $W_{0 \text{(GGA)}}$ are equilibrium interfacial distance and adhesion energy, respectively from reference \cite{Siegel2002}, where $d_{0 \text{(LDA)}}$ was calculated with LDA, and $W_{0 \text{(GGA)}}$ with PW91 after preceding relaxations with LDA.}
\begin{center}
\begin{tabular}{@{}*{12}{lccccccccccc}}
\hline
\hline
\multicolumn{1}{c}{} & \multicolumn{2}{c}{Before relaxation} & ~ & \multicolumn{2}{c}{After relaxation} & ~ & \multicolumn{2}{c}{Effect of relaxation} & ~ & \multicolumn{2}{c}{Reference \cite{Siegel2002}} \\
\cline{2-3} 
\cline{5-6} 
\cline{8-9}
\cline{11-12}
& $d_0$ & $W_0$ & ~ & $d_0$ & $W_0$ & ~ & $d_0$ & $W_0$ & ~ & $d_{0 \text{(LDA)}}$ & $W_{0 \text{(GGA)}}$\\
\hline
1Al FCC & 2.59 & 1.05 & ~ & 0.97 & 0.84 & ~ & $\downarrow$ & $\downarrow$ &~& 0.70 & 1.06\\
1Al HCP & 2.40 & 1.13 & ~ & 2.55 & 0.48 & ~ & $\uparrow$ & $\downarrow$ &~& 2.57 & 0.41\\
1Al OT & 2.24 & 1.30 & ~ & 2.29 & 0.70 & ~ & $\uparrow$ & $\downarrow$ &~& 1.62 & 0.84\\
\\
O FCC & 1.47 & 6.78 & ~ & 0.87 & 8.48 & ~ & $\downarrow$ & $\uparrow$ &~& 0.86 & 9.73\\
O HCP & 1.40 & 7.09 & ~ & 1.23 & 7.97 & ~ & $\downarrow$ & $\uparrow$ &~& 1.06 & 9.11\\
O OT & 1.73 & 7.41 & ~ & 1.72 & 8.05 & ~ & $\sim$ & $\uparrow$ &~& 2.00 & 8.75\\
\hline\hline
\end{tabular}
\end{center}
\label{relaxation}
\end{table*}

The distance at which the Al and \ox~slabs should be placed, in order to form the minimum-energy configuration, can be found by calculating the \textit{ideal work of adhesion} per unit area $W_{\text{Adh}}$ as a function of the interfacial distance $d$. $W_{\text{Adh}}$ is defined as

\begin{equation}\label{Wad_def} 
W_{\text{Adh}}=(E_{\text{Al}}+E_{\text{Al}_2\text{O}_3}-E_{\text{int}})/A , 
\end{equation}
where $E_{Al}$ and $E_{Al_2O_3}$ are the total energies of the isolated Al and {\ox} slabs, respectively, and $E_{int}$ is the total energy of the whole system. $A$ is the area of the interface. Thus, $W_{\text{Adh}}$ is equivalent to the energy needed to separate the two slabs infinitely from each other and has a positive sign for the bound system in equilibrium. Often in literature, the ideal work of adhesion is also referred as "the ideal work of separation" \cite{Finnis1996, iupac}. $-W_{\text{Adh}}$ can be interpreted as the \textit{adhesive binding energy} of the two films in the junction. 

A standard method for finding the optimal distance between the two slabs in the junction system, is to first calculate $-W_{\text{Adh}}$ at several distances with DFT. Next, a known analytic function is fit to the obtained points, and the equilibrium separation $d_0$ is identified where the minimum of the fit function occurs. The widely used analytic form for the adhesive binding energy is UBER (Universal Binding Energy Relation) \cite{Banerjea1988} 
\begin{equation}
W_{\text{UBER}}(d)=-W_0(1+d_s)\exp(-d_s),
\end{equation}
where $d_s=(d-d_0)/l$ is a scaling length, $l$ is a scaling parameter to be fitted, and $W_0$ is the fitted adhesion energy at the equilibrium separation $d_0$. UBER has been widely applied to various materials and interfaces, including Al/\ox~junctions \cite{Siegel2002}. However, to our knowledge, the accuracy of UBER specifically for Al/\ox~interfaces, has not been examined. Since the purpose of this work is to characterise Al/\ox~interfaces as components of tunnel devices, we consider the accurate determination of the metal-oxide separation particularly important. 

\begin{figure}[]
\centering
	\subfloat[]{	
		\includegraphics[width=0.23\textwidth]{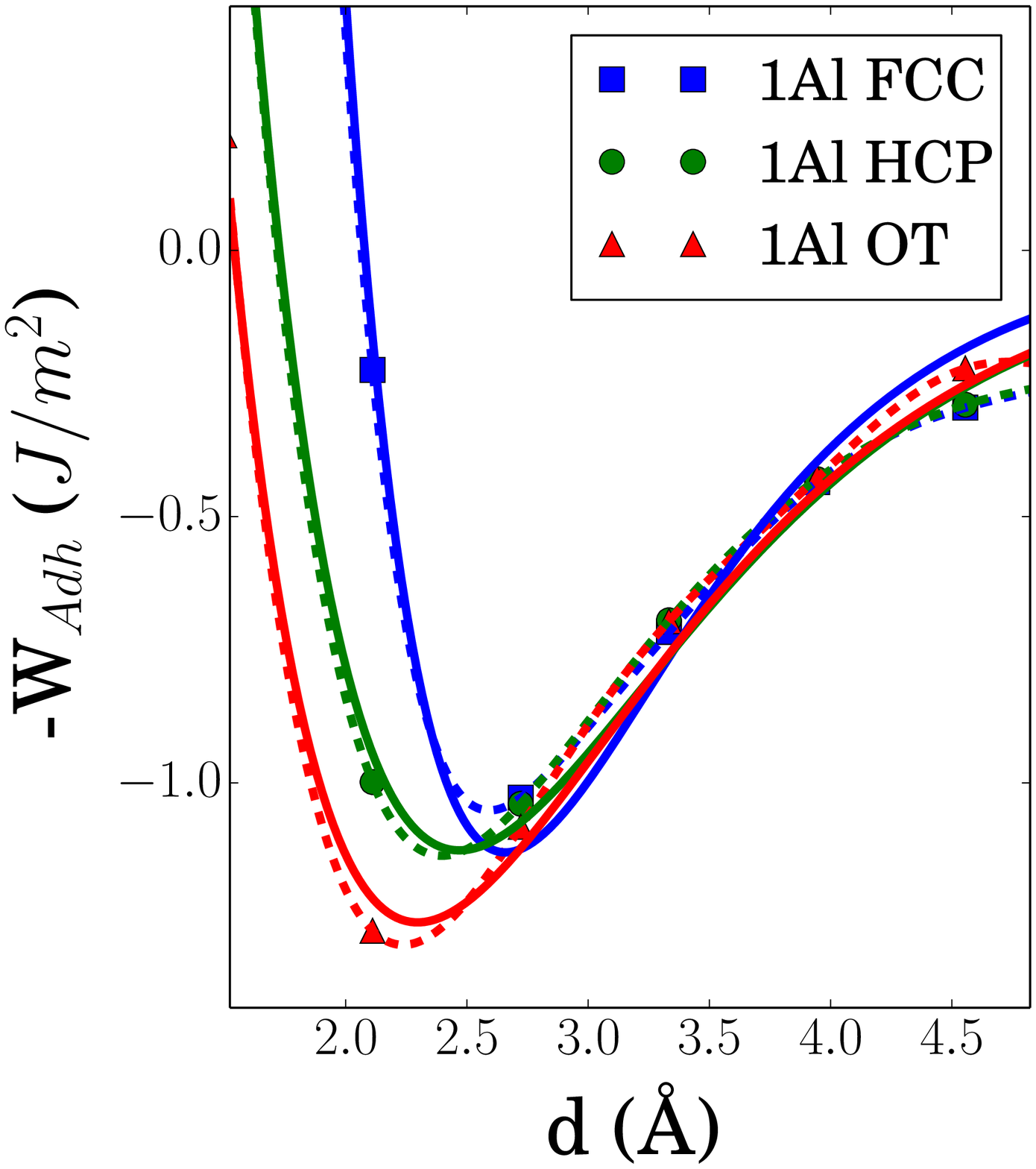} 
	}	
	\subfloat[]{
		\includegraphics[width=0.23\textwidth]{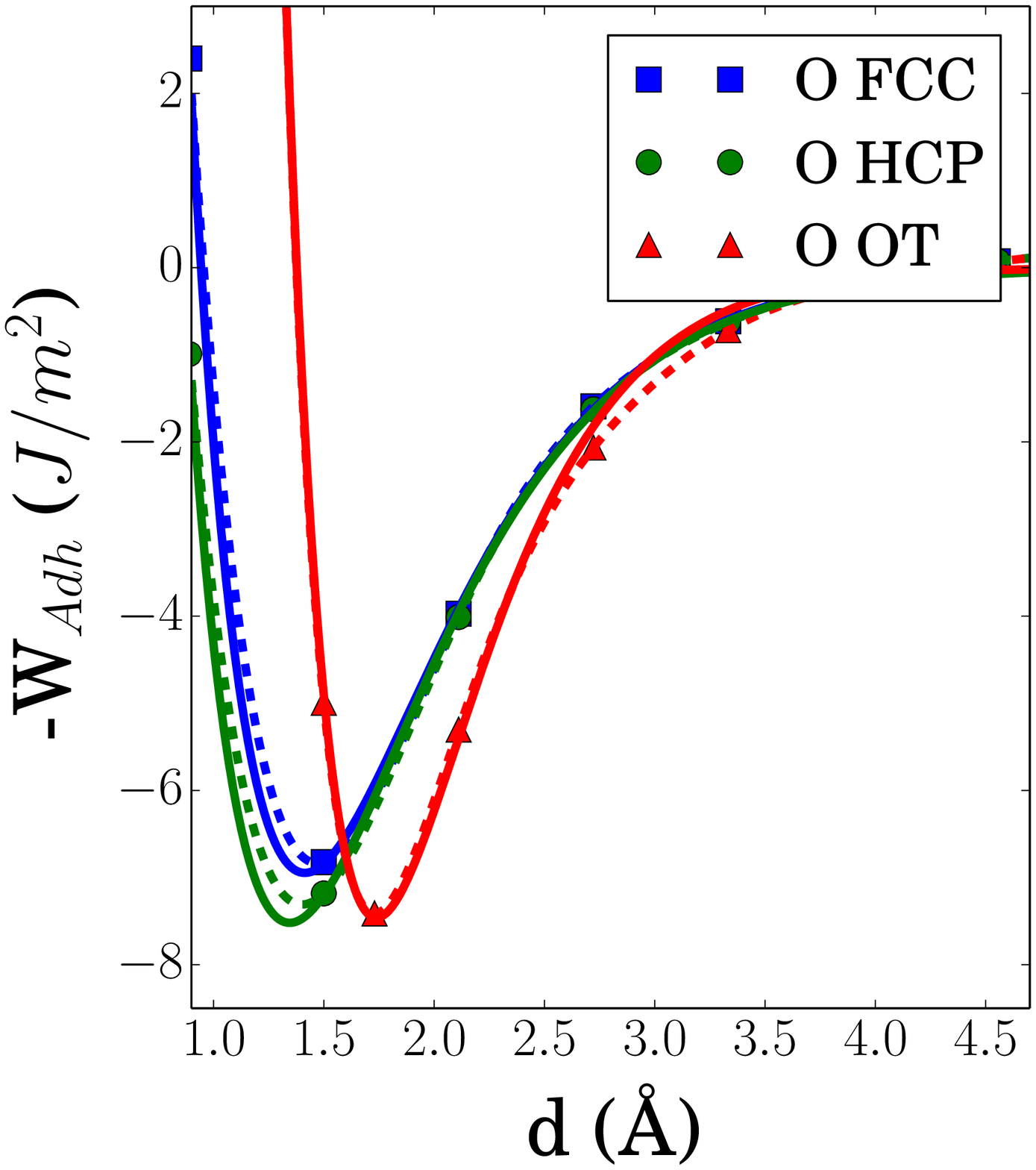} 
	}	
\caption{Adhesive binding energy as a function of the interfacial distance $d$ for the Al-terminated (a) and O-terminated (b) structures. The discrete data points are $-W_{\text{adh}}$ obtained from the DFT calculations. The solid lines are UBER fits to the DFT points. The dashed lines represent the 5th order spline fit.}
\label{Wad}
\end{figure}

UBER was developed to describe metallic and covalent bonds \cite{Rose1981, Rose1983, Banerjea1988} and might not be reliable for ionic solids. This dictates that we might expect the most reasonable description for the 1Al FCC structure, where the (unrelaxed) interface bonding is primarily formed via the Al-Al bonds. In contrast, the method might be particularly inaccurate in the case of the oxygen-terminated Al$_2$O$_3$ where ionic bonding dominates. Therefore, the applicability of UBER to our Al/\ox ~ systems has to be checked. For this purpose, we first predict $d_0$ and $W_{\text{Adh}}$ values using UBER, as well as using the 5th order spline fit to the DFT points (see figure \ref{Wad}). Next, we place the Al and \ox~layers at the distance $d_{0_{\text{UBER}}}$ ($d_{0_{\text{spline}}}$) and calculate the adhesion energies with DFT, one at the interfacial distance predicted by UBER ($W_\text{DFT}$(\duber)) and the other at the interfacial distance predicted by the 5th order spline  ($W_\text{DFT}$(\dspline)). The accuracies of the two fits are analysed in table \ref{fits}. When compared to the DFT results, the spline fit produces smaller errors than the UBER fit for all geometries. 

The spline interpolation results in 0-2.6 \% higher adhesion energies compared to those obtained using UBER. This means that the former predicts an interfacial distance at which adhesion is stronger, that is, it predicts the equilibrium  structure more accurately. On average, relative to the spline interpolation, UBER overestimates the interfacial distance for the Al-layer terminated structures by 2.8 \%, and underestimates it for the O-layer terminated structures by 2.6 \%. Nevertheless, UBER is usually a good enough approximation. However, as the accuracy of the interface geometry is crucial, we decide to use $d_0$ predicted by the spline fitting.

Finally, to set up the junction, we place the unrelaxed Al and \ox~slabs at the distances found from fitting and relax the entire structure. Relaxation results in the decrease of adhesion energy $W_0$  in all the Al-terminated interfaces, while in all the O-terminated cases, the effect is opposite (table \ref{relaxation}). The same trend has been observed for other metal/\ox~junctions as well \cite{Batyrev2001}. Even though structural relaxation leads to minimizing the total energy of a system, the same is not necessarily fulfilled for the adhesion energy. Whether $W_0$ increases or decreases following the relaxation, is defined by the \textit{change} in each term of equation \ref{Wad_def}. For example, increase in case of the O-terminated structures means that the net decrease in the total energies of the slabs is smaller than the decrease in the total energy of the junction, $\Delta(E_{\text{Al}})+\Delta(E_{\text{Al}_2\text{O}_3})<\Delta(E_{\text{int}})$. 

For the Al-terminated cases, the adhesion energy decreases in the following order, FCC, HCP, OT, while the interfacial distance $d_0$ increases in the same order. The trend agrees with the results reported in reference \cite{Siegel2002}. For the O-terminated structures $d_0$ has the same trend as that for the Al-terminated cases, while $W_0$ decreases in the following order: FCC, OT, HCP. This sequence differs from that obtained in reference \cite{Siegel2002}, and our predicted values for the adhesion energies are, in general, smaller. The FCC stackings exhibit the highest adhesion energies and smallest interfacial distances in agreement with reference \cite{Siegel2002}. 

\subsection{W$_\text{Adh}$ as a measure of stability?}

In the existing studies, detailed analyses of the interfaces with HCP and OT stacking are most often overlooked. Usually, the FCC-stacked interfaces are claimed to be the most stable structures and are examined as representatives of the junctions \cite{Siegel2002, Kim2013, Zhang2000}. One reason is the fact that these are the structures which exhibit the most dramatic relaxations of interfaces. Another reason is that according to the DFT calculations, the FCC stacking yields the highest ideal work of separation, interpreted as an indicator of the most stable configurations. Interestingly, in our previous study \cite{Koberidze2016}, by combining semi-classical methods and experimental data, we estimated that the most dominant geometry in the Al/\ox~junction should be 1Al OT followed by 1Al HCP. 
\begin{table*}[]
\caption{Binding energies $\Delta E$ of different Al/\ox~structures. Each system contains 45 atoms.} 
\label{delE}
\begin{center}
\begin{tabular}{l*{6}{c}}
\hline \hline
Geometry & 1Al FCC & 1Al HCP & 1Al OT & O FCC & O HCP & O OT \\
\hline
$\Delta E$ (eV) & 231.38 & 230.94 & 231.21 & 230.00 & 229.37 & 229.47 \\ 
\hline \hline
\end{tabular}
\end{center}
\end{table*}

Although the strongest interfacial adhesion should indeed mean the highest stability, the interpretation of $W_\text{Adh}$ for the relaxed interfaces should be considered more carefully: the definition of $W_\text{Adh}$ based on equation (\ref{Wad}) in case of the unrelaxed structures, is straightforward and unambiguous since the constituent slabs in the unrelaxed interfaces are structurally identical to the separate isolated slabs. In contrast, comparing the energy of the relaxed joint system to that of the isolated relaxed slabs, is vague, 
since Al or \ox, relaxed as parts of the junction, and respective independent slabs relaxed in vacuum, are not structurally identical. In addition to probable adjustments of the interplanar distances, the interface relaxation may cause changes in the local stoichiometry of \ox~and/or in the coordination of the Al atoms near the interface. Thus, in case of the relaxed interfaces, the definition of $W_\text{Adh}$ implies that if the junction was to be separated into the Al and \ox~parts, they would instantly adopt the geometry of the isolated slabs relaxed in vacuum. Such a definition neglects structural changes at the interface due to the relaxation. Moreover, whether in this way estimated $W_\text{Adh}$ is the measure of the adhesion between the immediate Al and \ox~surfaces, becomes ambiguous, since the termination or the start of either material in the interface region might no longer be well-defined in the relaxed junctions. Therefore, such a description leads to the inaccurate understanding of the interfacial adhesion strength, and consequently, of the stability of the junctions. To address the issue, we estimated the stability of the junctions by subtracting atomic energies from the total energy of the systems, analogously to calculating cohesive energies of the bulk materials: $\Delta E=N_{O}\cdot E_{O}+N_{Al} \cdot E_{Al}-E_{int}$, where $\Delta E$ is the binding energy of the system, $N_{O}$ and $N_{Al}$ are the number of O and Al atoms in the junctions, respectively. $E_{O}$ and $E_{Al}$ are the spin-polarized atom energies of O and Al, respectively and $E_{int}$ is the total energy of the system. Obtained results are presented in table \ref{delE}. While $\Delta E$ will not give information about the adhesion strengths between any two neighbouring layers, it provides an average estimate of the relative stabilities of the different junctions. Our results imply that, in general, 1Al-terminated interfaces are more stable compared to the O-terminated cases. More importantly, 1Al OT is virtually as stable as 1Al FCC, supporting our prediction in the previous work. 

\subsection{Atomic relaxations at the Al/$\alpha$-A\lowercase{l}$_2$O$_3$ interfaces}
As mentioned in the introduction, atomic structure of the interface has a  critical impact on the behaviour of the MIM devices, especially, on the functionality of the oxide barrier. The barrier performance is traditionally modelled using rectangular \cite{Simmons1963, Arakawa2006, Jung2009, Mistry2017}, and sometimes trapezoidal \cite{Bokes_conductance, Koberidze2016} tunnel barrier models. Detailed characterization of the interfacial structure is crucial to identify the factors affecting the tunnel barrier parameters (width, height, abruptness), and to associate them with the geometrical patterns of the interface. Below, we will describe the re-established bond lengths and the atomic composition in the interface region following the relaxations. 

\begin{figure*}[ht]
	\subfloat[1Al FCC]{	
		\includegraphics[trim={0 0 0 1.0cm}, clip, width=0.325\textwidth]{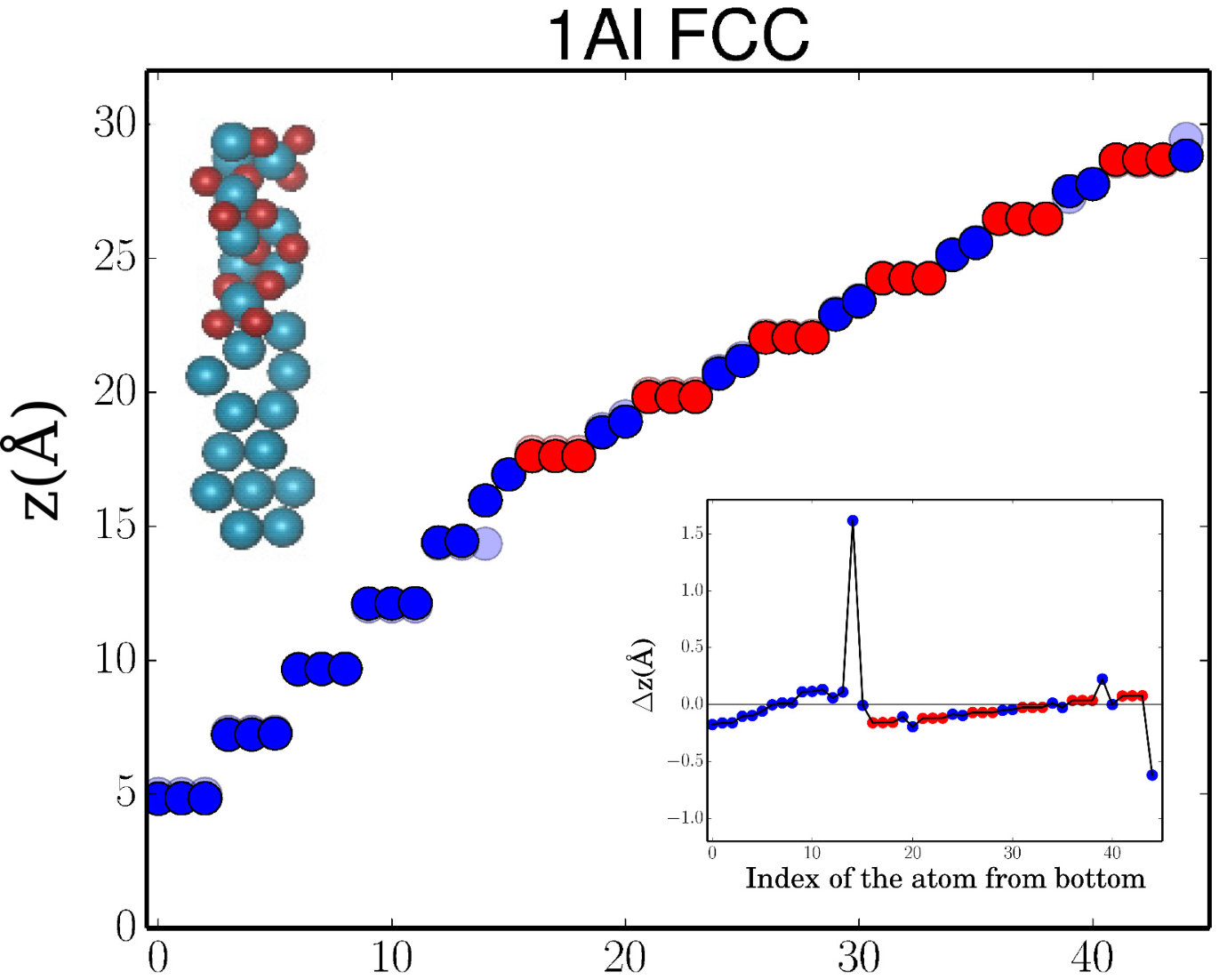} 
	}	
	\subfloat[1Al HCP]{
		\includegraphics[trim={0 0 0 1.0cm}, clip, width=0.3\textwidth]{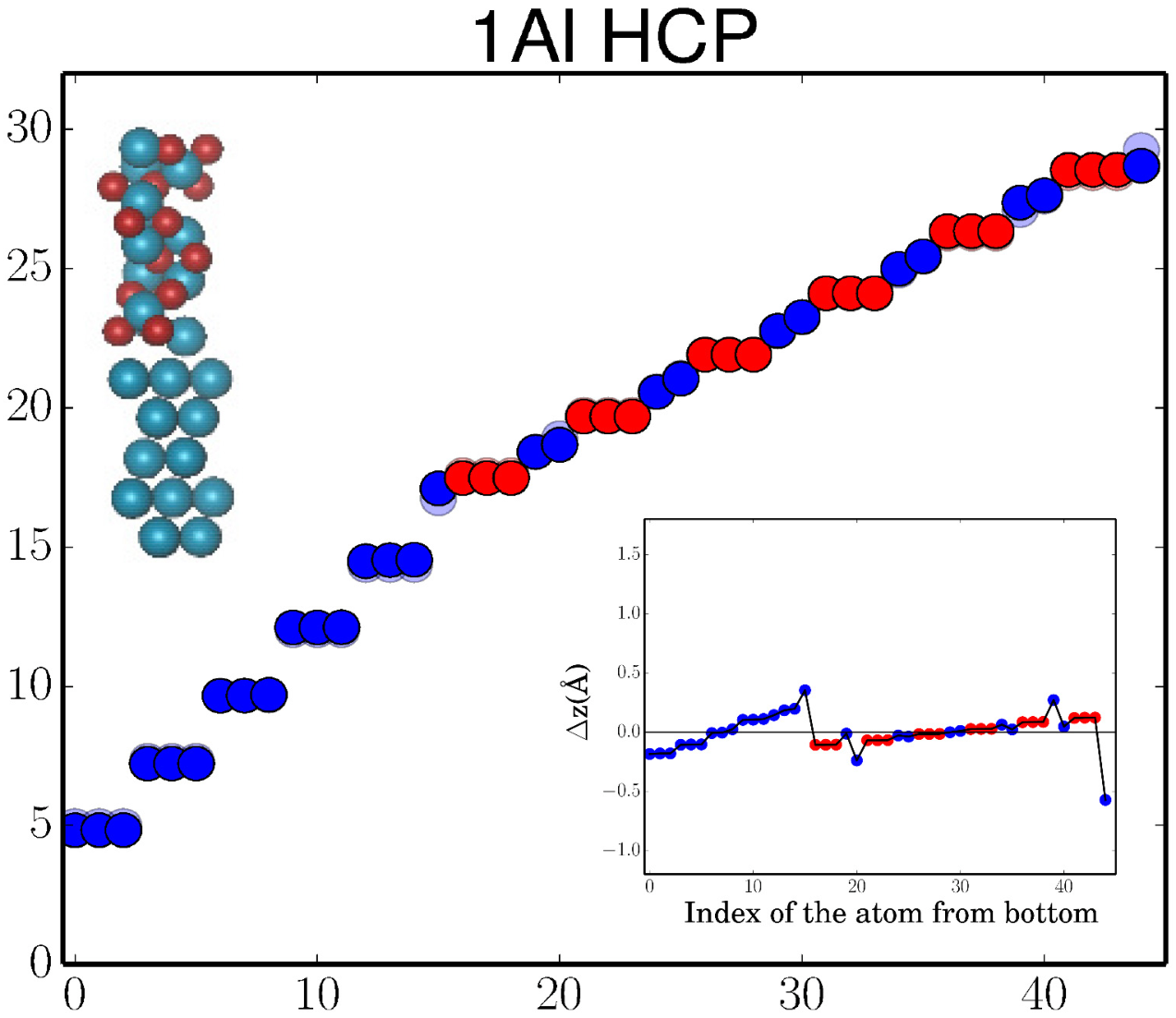}
	}	
		\subfloat[1Al OT]{
		\includegraphics[trim={0 0 0 1.0cm}, clip, width=0.3\textwidth]{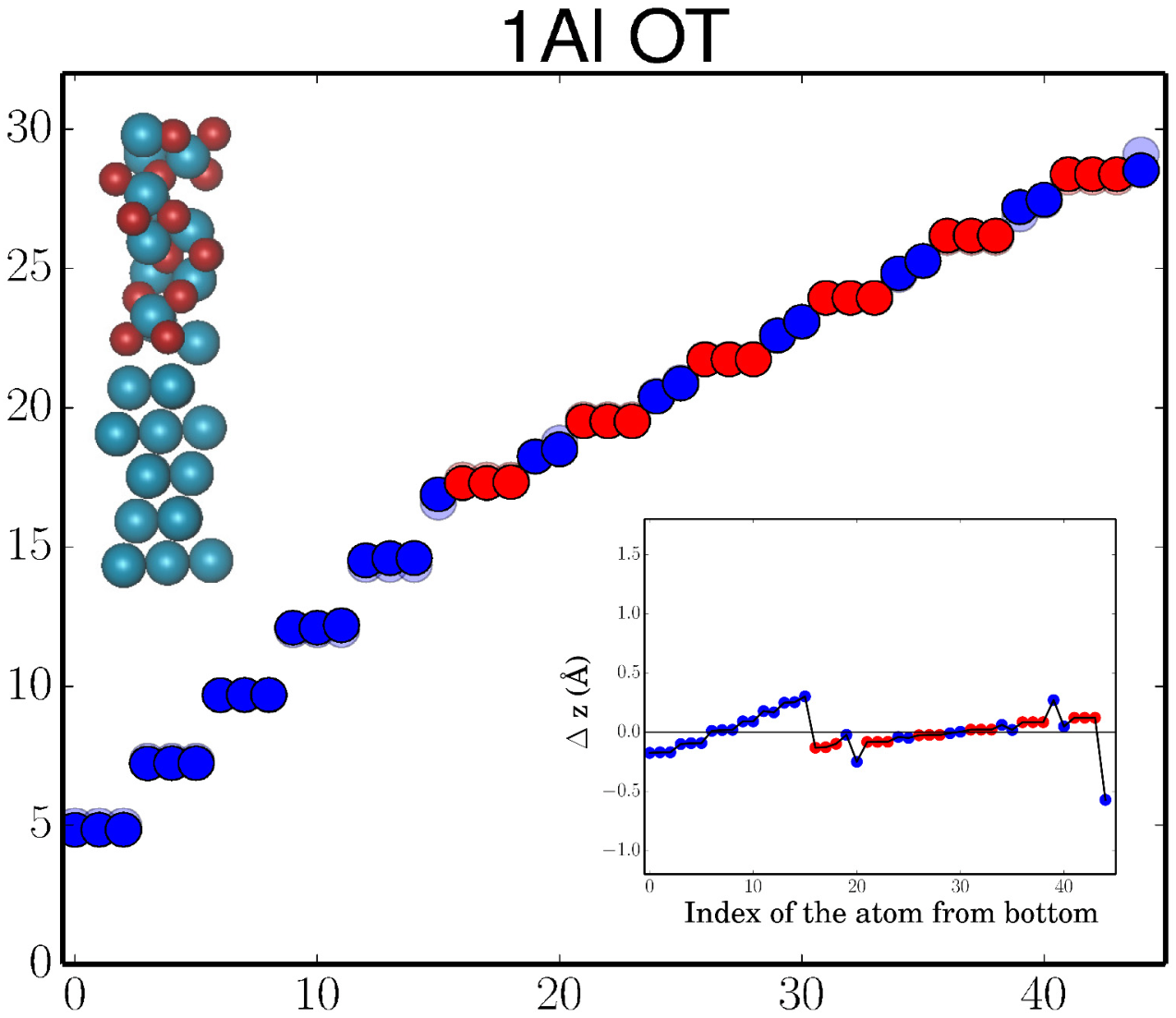} 
	}	\\
		\subfloat[O FCC]{
		\includegraphics[trim={0 0 0 1.0cm}, clip, width=0.325\textwidth]{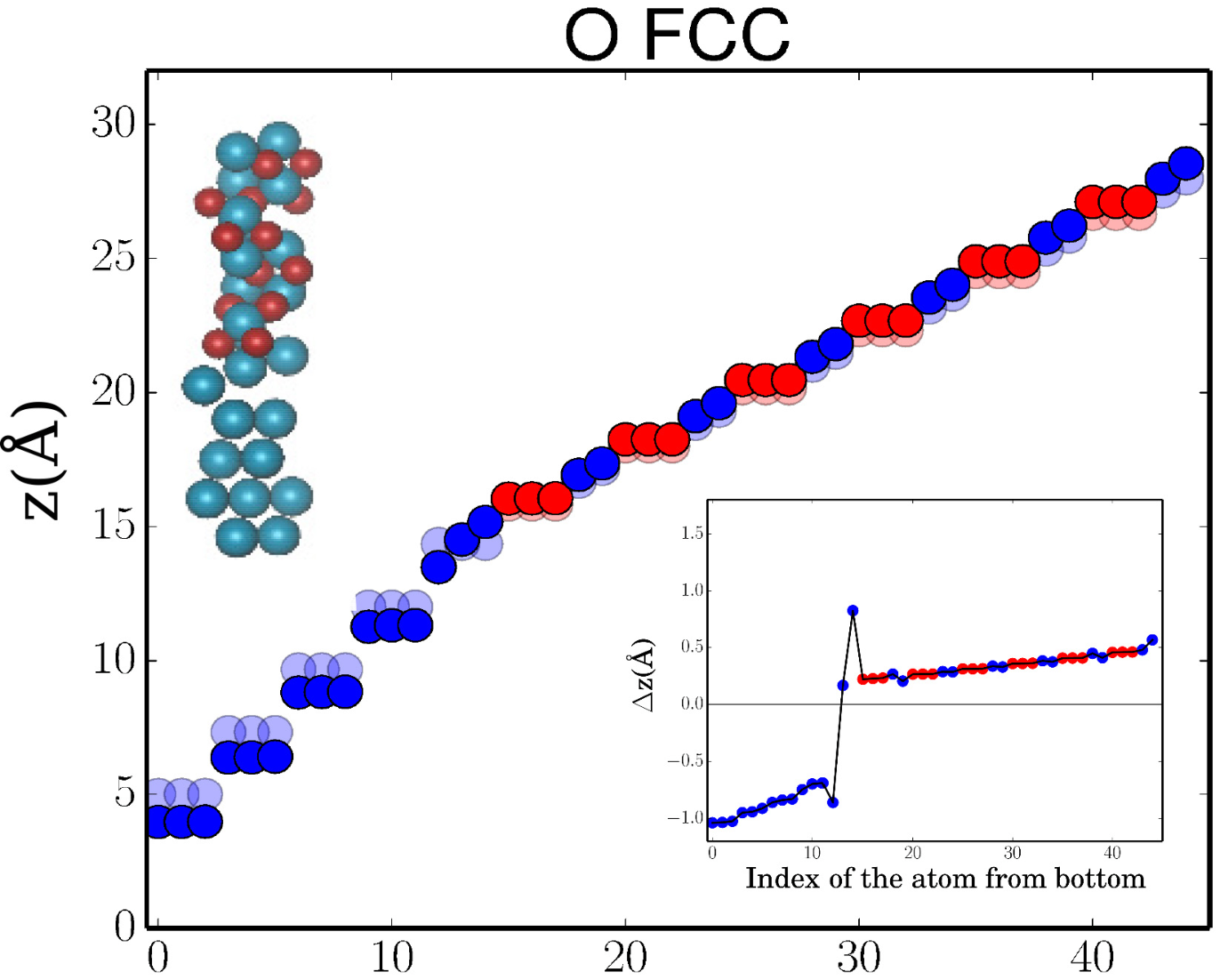}
	}
		\subfloat[O HCP]{
		\includegraphics[trim={0 0 0 1.0cm}, clip, width=0.3\textwidth]{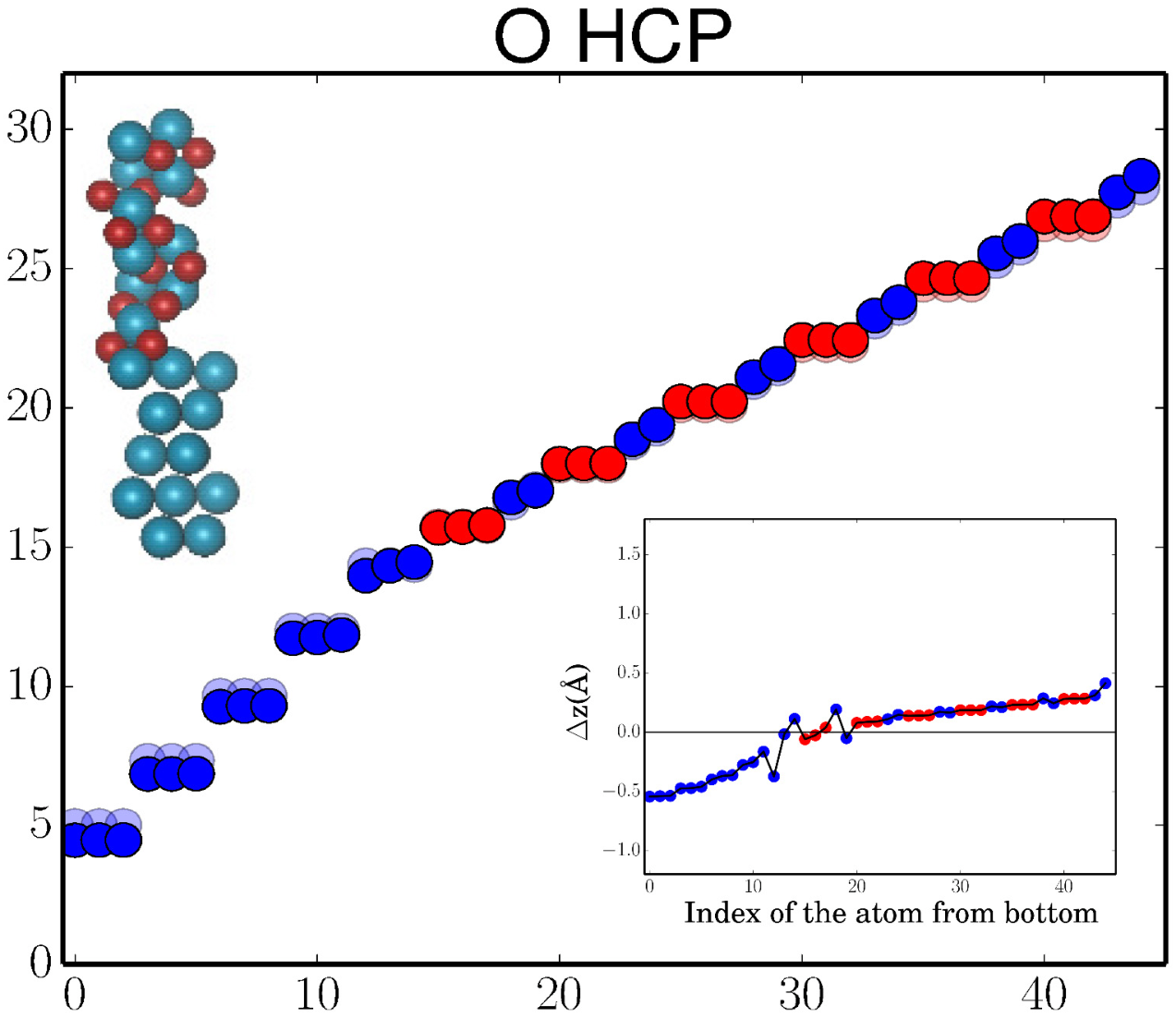}
	}
		\subfloat[O OT]{
		\includegraphics[trim={0 0 0 1.0cm}, clip, width=0.3\textwidth]{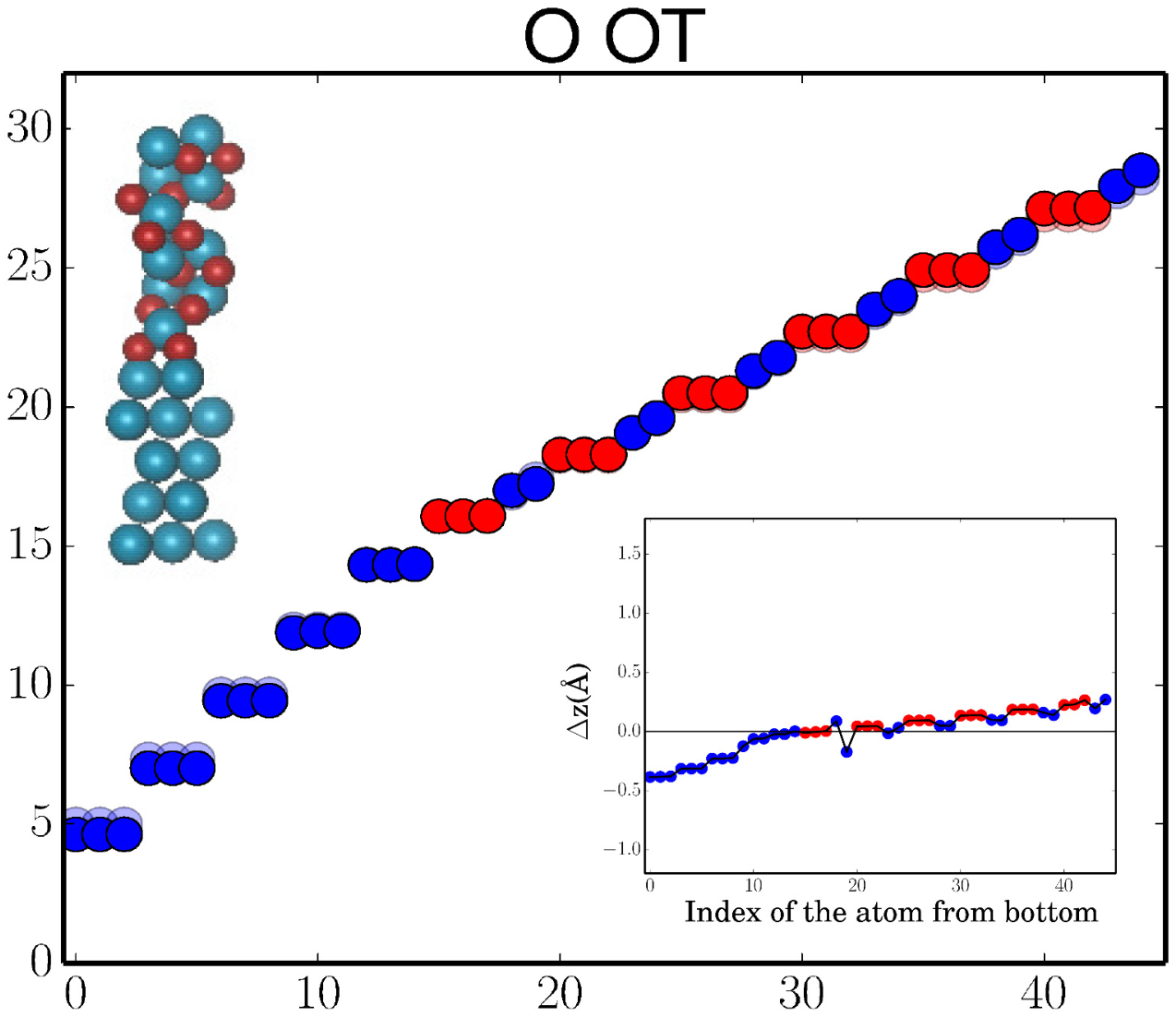}
	}
\caption{Atomic relaxation at the six interfaces along the $z$ coordinate. (a)-(c) illustrate the Al-terminated cases, and (d)-(f) - the O-terminated cases. The blue circles represent Al atoms and the red ones O atoms. The light blue and red circles show the unrelaxed positions of the Al and O atoms, respectively. The insets show the changes in the $z$ coordinates for each atom with respect to its unrelaxed position. The images in the upper left corners of the panels are the relaxed geometries from the simulations.}
\label{Z}
\end{figure*}

Relaxation-induced atomic rearrangements at the interfaces vary with the oxide termination and the stacking sequence of the joined surfaces (figure \ref{Z}). Displacements are significant along the $z$ direction. The center-of-mass translation occurs both in the metal and the oxide parts.

Relaxation of the interfacial Al atoms belonging to the metal films tend to cause structural changes in the interface regions. The effect is particularly notable for the 1Al FCC and O FCC junctions. In 1Al FCC, one of the Al atoms from the surface layer of the metal film 
is pulled towards the first oxide layer at the interface complementing the stoichiometry of the \ox~unit. The nearest neighbour to the displaced Al atom becomes O from the first oxygen layer of the oxide, instead of the Al atom from the metal film. The new nearest neighbour bond length amounts to 2.18 \AA, which exceeds 1.97 \AA, a long Al-O nearest neighbour bond distance in the pristine bulk oxide. Thus, the initially Al-terminated oxide film becomes 2Al-terminated, and an Al vacancy is introduced in the original surface layer of the metal film (figure \ref{Z} (a)). 

The O FCC interface undergoes an analogous relaxation as its Al-terminated counterpart (figure \ref{Z} (d)). However, in this case, complementing the \ox~stoichiometry requires pulling two atoms from the surface of the metal film. The resulting nearest neighbour distances for the two displaced Al atoms are 2.11 \AA~ and 1.86 \AA. The corresponding distances in the bulk oxide are 1.97 \AA~ and 1.86 \AA~ for the long and the short Al-O bond, respectively. The initially O-terminated oxide film becomes 2Al-terminated, similarly to the 1Al FCC junction. This creates a divacancy on the surface of the metal film. 

In the O HCP system, relaxation distorts the metal surface atoms similarly to those in the O FCC junction, though with only slight deviations from their initial positions. Unlike for the O FCC junction, the \ox~ stoichiometry in the O HCP geometry is not restored due to the unfavourable positions of the interface Al and O layers with respect to each other. 

In the remaining three junctions, 1Al HCP, 1Al OT and O OT, rearrangements of the atoms near the interface, belonging to the metal film, are relatively small. The nearest neighbour Al-O distances in the 1Al OT and 1Al HCP geometries, in the interface region, are smallest among all the six interfaces and equal to 1.73 \AA~ for both structures. The next shortest Al-O bond length is at the interface of the O OT structure amounting to 1.75 \AA. Both the values are smaller than the shortest Al-O distance in the bulk oxide, which is 1.86 \AA~ as mentioned above.

In a recent study \cite{Zeng2016}, nearest neighbour distances in the  Al/AlOx/Al junction were extracted from the Nanobeam Electron Diffraction (NBED) data. Based on the pair distribution function (PDF) analysis, the most occurrent Al-O distances were found to be equal to 1.76 \AA. Less prevalent Al-O bond lengths of 2.27 \AA~ were also observed and assumed to correspond to the interfacial Al-O bonds. The first value is close to 1.73-1.75 \AA, which we found to be present at the interfaces of the 1Al HCP, 1Al OT and O OT geometries. The second value is close to 2.18 \AA~ and 2.11 \AA~ which we observed at the interfaces of the 1Al FCC and O FCC geometries, respectively. This implies that the shortened Al-O distances at the interfaces (compared to the bulk value), have contributed to reducing the average value of the shortest Al-O distances in the oxide volume probed by the measurement. The effect could be significant because of the small thickness of the oxide. 

In summary, while relaxing the interface, the oxide film tries to extend its crystalline structure and fills its vacant Al sites with the Al atoms from the metal film within the available space for relaxation, as also observed in earlier studies \cite{Siegel2002, Liu2014}. As a result, O FCC and 1Al FCC interfaces become 2Al-terminated, while 1Al HCP and 1Al OT maintain 1Al-terminated structure. Only O HCP and O OT remain O-terminated. However, we will continue labelling the interfaces by their initial unrelaxed geometries. 

Figures \ref{Z} and \ref{interfaces} show that the transition region between the metal and the oxide films may be either relatively abrupt or span several atomic layers at the interface, and it is primarily composed of the Al atoms displaced from the surface of the metal film. The short-range roughness on the surface of the metal substrate in Al/\ox~junctions with the height of one or two Al(111) interplanar spacings has also been observed experimentally \cite{Timsit1985}. Moreover, experimental studies have shown that the surface roughness of the metal film strongly influences the conductive properties of the MIM junctions \cite{Alimardani2012, Kohlstedt1996, Shen2012}. In our previous work, we found that 1Al FCC was characterised with the smallest barrier height, followed by O FCC and O HCP. These are the structures where the metal surfaces are most distorted. Thus, our results  suggest that deviations from coplanarity of the metal surface layers has a critical role in the formation of the tunnel barrier, and should lower its height. This can be explained by the fact that a disordered and scarce distribution of Al atoms near the interface causes a smeared and reduced charge density in this area compared to the bulk metal. The more scarce the Al atom distribution is, the smaller the charge density, and the higher the electrostatic potential at the interface. The affected electrostatic potential changes the relative positions of the Fermi level of the metal and the conduction band edge of the oxide, which defines the height of the tunnel barrier. However, the positions of the energy levels depend not only on the spacial configuration of the atoms but also on the possible charge transfer at the interface. Explaining the band alignment needs a more detailed investigation of the charge densities and the potentials for different interface geometries. In this paper, we focus solely on the geometrical properties of the junctions.

\subsection{Interlayer relaxations in A\lowercase{l}/A\lowercase{l}$_2$O$_3$ systems}

Besides introducing irregularities at the surfaces of the metal films, relaxation effectively changes the thicknesses of the insulating layers in metal-oxide systems. This is due to acquirement of Al atoms from the metal surface and/or due to the changes in the interplanar distances in the oxide. The variation in the insulator thickness throughout a MIM junction is a commonly observed phenomenon in experiments. It highly affects the operation of MIM devices, since the tunnelling probability depends exponentially on the barrier width. Even though the most probable reason for the thickness variations in experiments could be the varying number of layers, changes in thicknesses due to interplanar relaxations might further enhance the non-uniformity within a sample. This means that the thickness of an oxide composed of a defined number of layers, might still vary along the junction because of the differences in the local geometries in different regions of the interface.

\begin{figure*}[ht]
	\subfloat[]{	
		\includegraphics[width=0.4\textwidth]{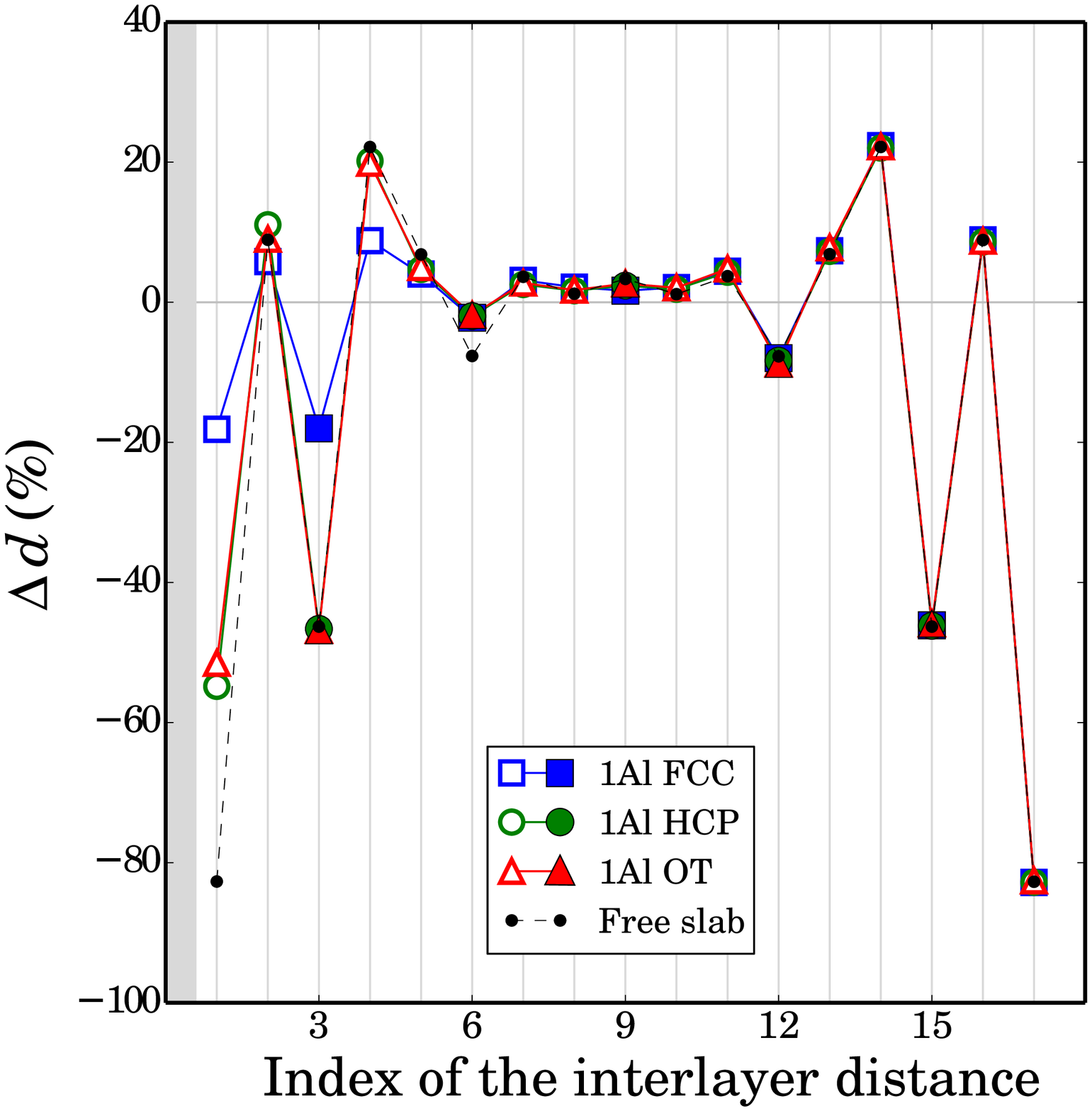} 
	}	
	\subfloat[]{	
		\includegraphics[width=0.4\textwidth]{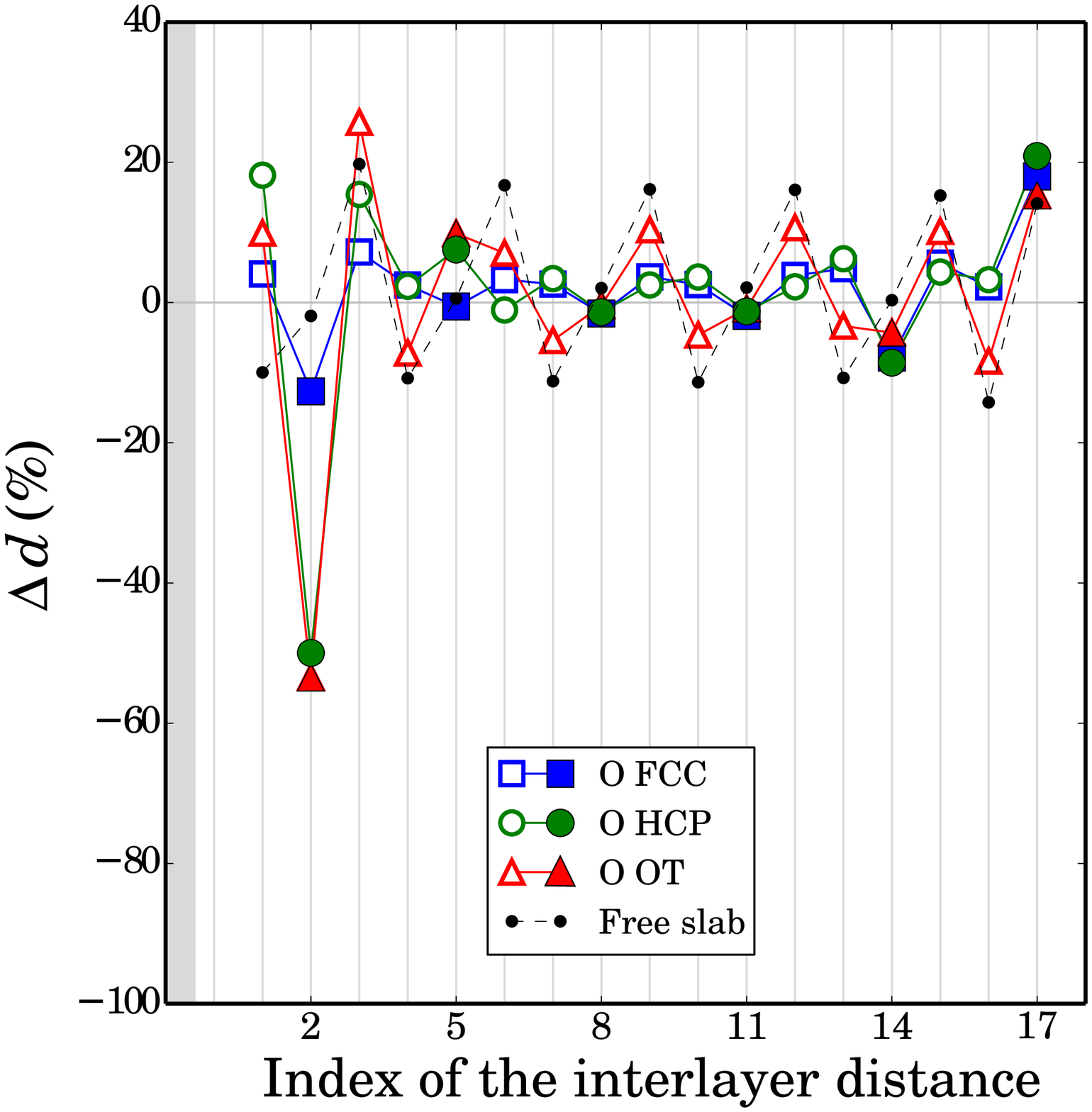} 
		}\\
	\subfloat[]{	
		\includegraphics[width=0.4\textwidth]{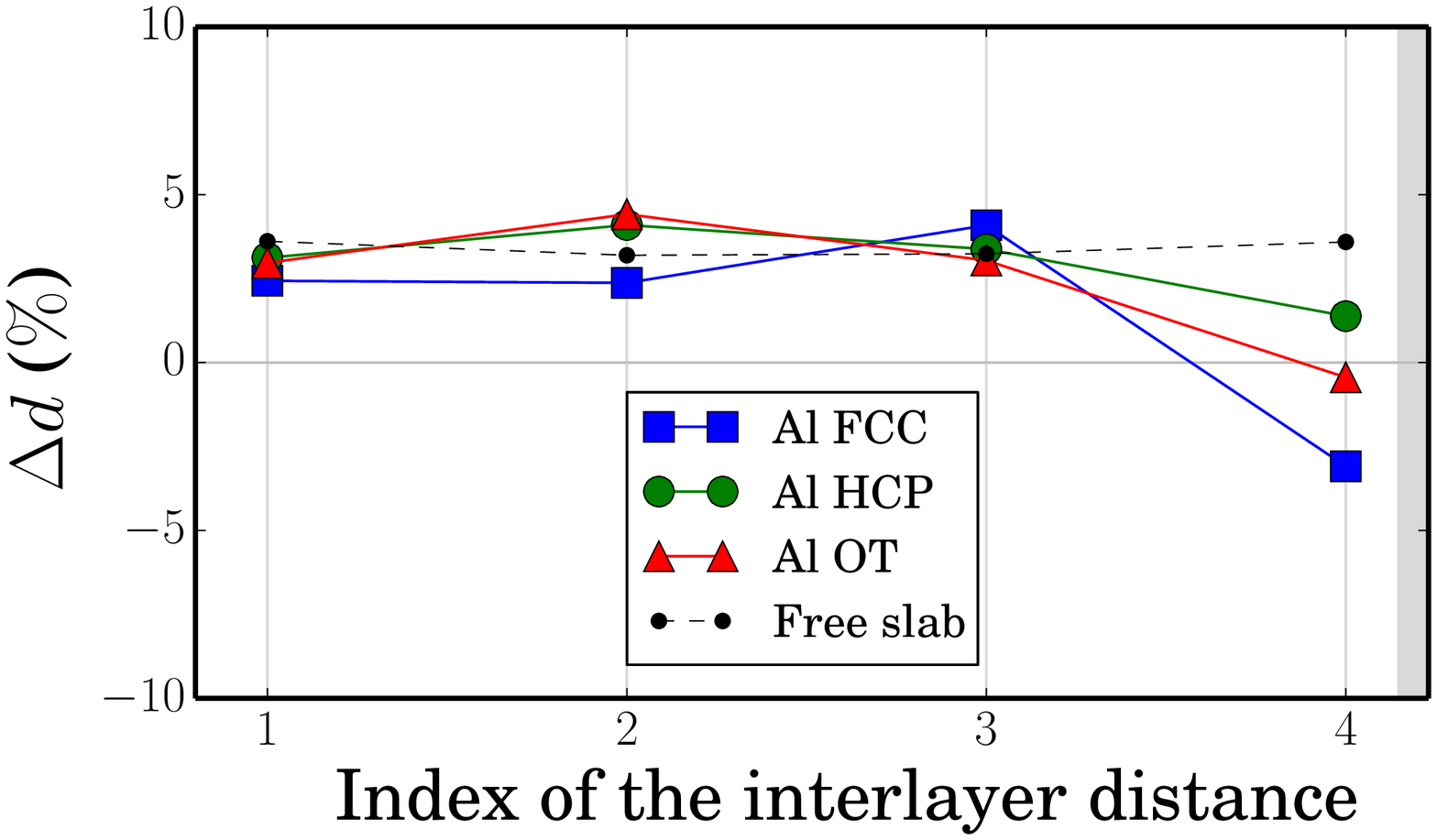}
		}
	\subfloat[]{	
		\includegraphics[width=0.4\textwidth]{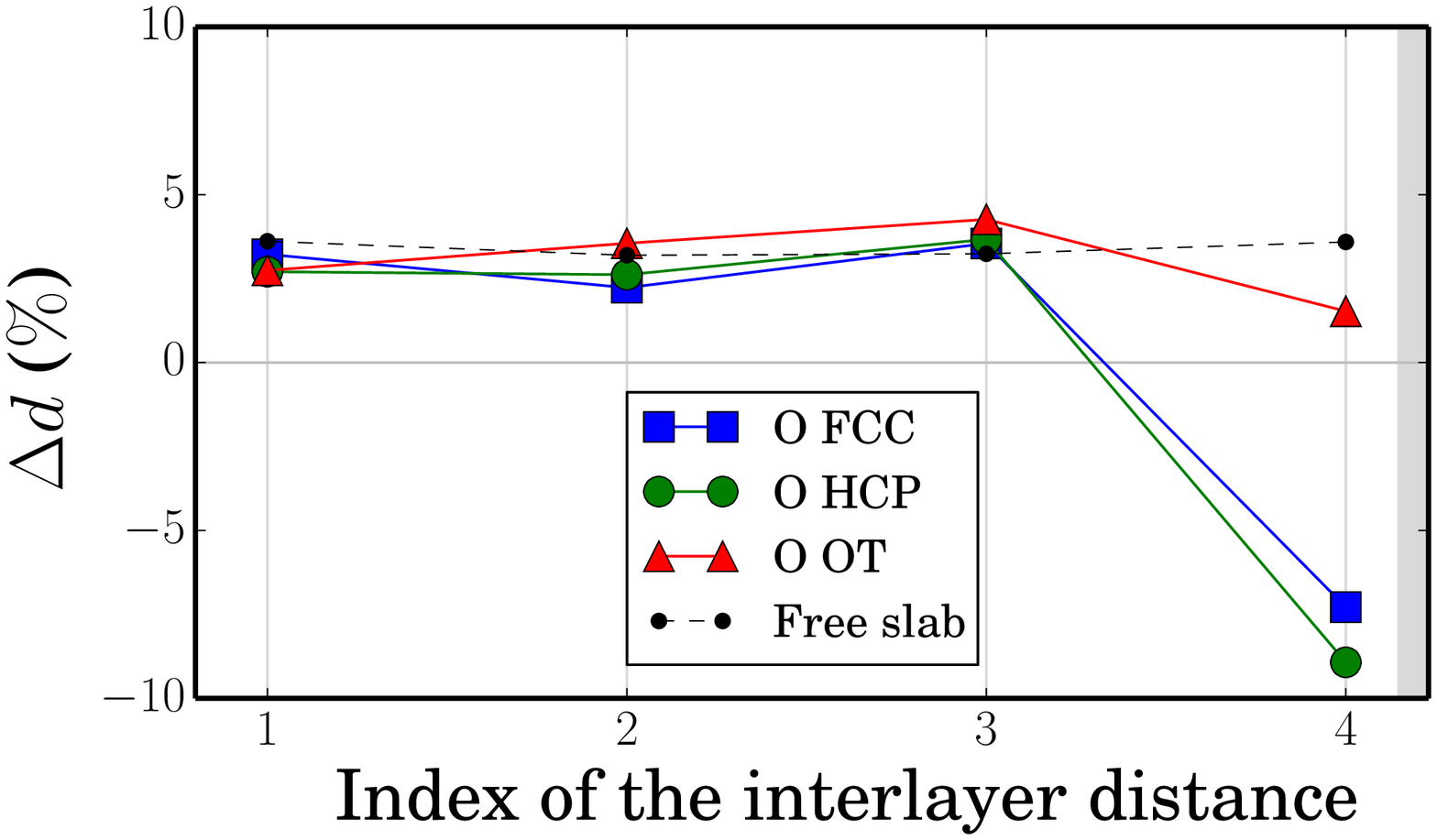}
		}	
\caption{Interplanar space relaxation $\Delta d$ relative to the bulk interlayer distances in the metal and the oxide films of Al/\ox~junctions. (a) and (b) - \ox~part in the O-terminated junctions, (c) and (d) - Al metal part in the Al-terminated junctions. Shaded regions at the edges of the plots show the positions of the interfaces. Filled and open data points correspond to the Al-Al and Al-O interplanar distances, respectively.}
\label{dz_int2}
\end{figure*}

To find out how far the influence of the interfacial geometry extends in the junction, we examine the relaxations of the interlayer distances beyond the interfaces. Figure \ref{dz_int2} shows the interlayer relaxations $\Delta d$ relative to the bulk interlayer distances in the Al and \ox~parts of the studied systems. Al-Al interplanar distances are denoted with the filled markers and Al-O interplanar distances with the open ones. The numbers of layers correspond to those in the Al or \ox~slabs before the relaxation of the total system. Thus, additional layers appearing due to the interface relaxation are omitted in order to enable comparison to the pristine metal or oxide surfaces.

Figure \ref{dz_int2} (a) presents results for the oxide films of the junctions with the Al-terminated  \ox~(the solid lines) in comparison with the relaxed isolated Al-terminated \ox~slab (the dashed line). The different stacking sequences and the interfacial distances affect interlayer relaxation of the oxide up to the fourth interlayer (index 4 on the horizontal axis) or approximately 2.6 \AA~for the HCP and OT stackings, and 2.9 \AA~ for the FCC stacking, after which $\Delta d$ distances do not differ notably regardless of the interface type. As a general trend, similarly to the pristine \ox~slab, Al-Al vertical distances are contracted compared to the bulk and Al-O vertical distances are expanded, except the first Al-O interlayers (index 1), which experience significant, although smaller contractions compared to the clean surface. This can be attributed to the presence of the Al-substrate. The second Al-Al interlayer distances (index 6) are also constrained closer to the bulk value than those of the pristine slab. Overall, 1Al HCP and 1Al OT follow relatively closely to the behaviour of $\Delta d$ for the isolated \ox~slab. These are the two junctions, which exhibited the least changes in the atomic positions during the interface relaxation, and which are found to experience the biggest contraction of the Al-O interplanar spacings near the interface. In contrast, the oxide part in the 1Al FCC system relaxes towards the bulk structure rather than towards the free surface. 

The O-terminated interfaces display a different behaviour (figure \ref{dz_int2} (b)). In the O FCC and O HCP junctions, interplanar distances are affected by the interface relaxation up to the index 6, or 4.4 \AA~from the interface. Starting from the index 7, $\Delta d$ values of the two junctions closely follow each other and fluctuate around the bulk values. Virtually all the Al-O interlayer distances in O FCC and O HCP junctions experience expansions of varying magnitude, particularly near the interfaces. In the case of the O OT junction, interplanar relaxation qualitatively follows the other two cases only up to the index 3, after which every other Al-O interlayer distance is contracted compared to the bulk value. The behaviour is also observed in the free oxide slab and is absent in the O FCC and O HCP cases. In all the O-terminated junctions, the first O-Al interplanar distance is expanded, contrary to the clean surface. Similarly to the Al-terminated cases, even though the O FCC structure exhibits the most pronounced atomic rearrangement at the interface, the interplanar distances in the oxide part of the junction relax closer to the bulk values. The O HCP junction recovers its bulk-like structure starting from the forth interplanar distance, while all the $\Delta d$ values of O OT evolve closest to the clean surface. 

Interplanar relaxations in the metal parts are illustrated in figures \ref{dz_int2} (c) and (d) for the Al- and O-terminated junctions, respectively. In all the geometries, the first three interplanar spacings experience slight expansions (\textless 5\%), similarly to the pristine  slab. This is due to the lateral compressive strain on the Al metal film needed to match the lattice constant of the oxide. The effect was verified for Au(001) in Au(001)/Fe(001) junctions in reference \cite{Benoit2012}. It was also shown that the changes in the interlayer distances closer to the interface (index 4 in our case) is not an artefact of the compressive strain on the metal, which is also visible in our results, since $\Delta d$ for the named interlayers differs in magnitude as well as in sign for different junctions. We observe relatively small expansions (\textless 3 $\%$) of the fourth interlayers for the 1Al HCP and O OT structures, and stronger contractions for the O HCP and O OT structures (about 7-9 $\%$). In all the junctions, the fourth interlayer distance of the Al metal part is smaller than in the independent (compressed) Al slab. Nevertheless, these interplanar relaxations are significantly smaller than those observed in the oxide parts, even though the Al parts experience stronger atomic displacements at the interface compared to the oxide parts as described in the previous section. 

To summarize, an overall impact of interplanar relaxations on the thicknesses of the oxide is either contraction or expansion depending on the interface geometry. In O-terminated junctions, the total thicknesses of the first seven layers of the oxide (up to the third O layer) are slightly expanded (by 2 \%) compared to the bulk case. Thus, interplanar distances fluctuate in a compensating manner to maintain the net width. In 1Al HCP and 1Al OT structures, the first five layers (up to the second O layer) experience net contractions by 13 \%, and in 1Al FCC by 3.3 \%. In the 1Al FCC and O FCC junctions, the oxide interlayers expand up to the second O layer by 30 \% and 37 \%, respectively, relative to the bulk. This is due to acquirement of Al atoms from the metal film by the oxide.

\section{Summary}
We have studied in detail the structures of the six Al/\ox~interfaces and their roles in formation of the tunnel barrier profiles using the first principles DFT modelling. We have checked the accuracy of the commonly used methods in existing  works on Al/\ox~interfaces. These are the exchange-correlation functionals, the universal binding energy relation and the ideal work of separation. We have found that PBE is the most relevant functional for describing Al/\ox~systems, and that the higher-order spline interpolation outperforms the UBER fit for finding equilibrium interfacial distances and adhesion energies. In addition, we have challenged the interpretation of the ideal work of separation as a measure of the junction stability and have shown that it might be misleading in predicting the optimal structures. Instead, total binding energies of the studied junction models could be more relevant for estimating the overall stability of the systems. 

After carefully obtaining the stable structures, we have thoroughly characterised the atomic and interplanar relaxations in the junctions. Interplanar relaxations span up to the fifth layer for the Al-terminated structures and up to the seventh layer for the O-terminated structures. Beyond these layers, interplanar relaxations are independent of the interface geometry. The initially Al-terminated oxides tend to decrease the thickness, while the O-terminated ones experience expansion. 

By examining the systems with identical numbers of layers, we have identified the three main contributions that have a significant effect on the electron tunnel barrier parameters: i) interplanar relaxations in the oxide beyond the immediate interface, which contribute to the variation of the barrier width ii) irregularities on the metal film surface, which lower the barrier height and expand the metal-oxide transition region iii) extension of the oxide layers by adopting Al atoms from the metal surface, which expands the width of the barrier. 

The gained information provides further insight into the correlation between the detailed atomic structure and the tunnel barrier parameters, which determines the performance and the stability of MIM devices.

\begin{acknowledgments}
We acknowledge the computational resources provided by the Aalto Science-IT project. This work was supported by the Academy of Finland through its Centres of Excellence Programme (2012-2017) under Project No. 251748. 
\\

\end{acknowledgments}

\bibliography{references}

\end{document}